\documentclass[useAMS,usenatbib]{mn2e}
\usepackage{amsmath,graphicx,fleqn,txfonts}

\title{Dynamical insight into dark-matter haloes}

\author[Walter~Dehnen \& Dean~E.~McLaughlin]
{
  Walter Dehnen%
\thanks{Email: walter.dehnen, dean.mclaughlin@astro.le.ac.uk}
  and Dean E.~McLaughlin$^{\star}$\\
  Department of Physics \& Astronomy,
  University of Leicester,
  Leicester, LE1~7RH
}

\date{Accepted .
      Received ;
      }
 
\pagerange{\pageref{firstpage}--\pageref{lastpage}}
\pubyear{2005}
 
\begin{document}

\maketitle
\label{firstpage}

\begin{abstract}
  We investigate, using the spherical Jeans equation, self-gravitating dynamical
  equilibria satisfying a relation $\rho/\sigma_r^3\propto r^{-\alpha}$, which
  holds for simulated dark-matter haloes over their whole resolved radial range.
  Considering first the case of velocity isotropy, we find that this problem has
  only one solution for which the density profile is not truncated or otherwise
  unrealistic. This solution occurs only for a critical value of
  $\alpha_{\mathrm{crit}} =35/18 =1.9\overline{4}$, which is consistent with the
  empirical value of $1.9\pm0.05$. We extend our analysis in two ways: first we
  introduce a parameter $\epsilon$ to allow for a more general relation
  $\rho/\sigma_r^\epsilon\propto r^{-\alpha}$; second we consider velocity
  anisotropy parameterised by Binney's $\beta(r) \equiv 1- \sigma_\theta^2 /
  \sigma_r^2$. If we assume $\beta$ to be linearly related to the logarithmic
  density slope $\gamma(r)\equiv-\mathrm{d}\ln \rho/\mathrm{d}\ln r$, which is
  in agreement with simulations, the problem remains analytically tractable and
  is equivalent to the simpler isotropic case: there exists only one physical
  solution, which occurs at a critical $\alpha$ value.  Remarkably, this value
  of $\alpha$ and the density and velocity-dispersion profiles depend only on
  $\epsilon$ and the value $\beta_0\equiv\beta(r=0)$, but not on the value
  $\beta_\infty\equiv\beta(r\to\infty)$ (or, equivalently, the slope
  $\mathrm{d}\beta / \mathrm{d}\gamma$ of the adopted linear $\beta$-$\gamma$
  relation). For $\epsilon=3$, $\alpha_{\mathrm{crit}} =35/18-2\beta_0/9$ and
  the resulting density profile is fully analytic (as are the velocity
  dispersion and circular speed) with an inner cusp $\rho\propto
  r^{-(7+10\beta_0)/9}$ and a very smooth transition to a steeper outer
  power-law asymptote. These models are in excellent agreement with the density,
  velocity-dispersion and anisotropy profiles of simulated dark-matter haloes
  over their full resolved radial range. If $\epsilon=3$ is a universal
  constant, some scatter in $\beta_0\approx0$ may account for some diversity in
  the density profiles, provided a relation $\rho/\sigma_r^3\propto
  r^{-\alpha_{\mathrm{crit}}}$ always holds.
\end{abstract}
\begin{keywords}
  stellar dynamics -- 
  methods: analytical -- 
  galaxies: haloes --
  galaxies: structure
\end{keywords}

\section{Introduction}
\label{sec:intro}

It has long been recognised that $N$-body studies of large-scale structure
formation in cold dark matter (CDM) cosmologies produce dark-matter haloes whose
density profiles are remarkably similar in shape over a wide range of halo
virial mass \citep[e.g.][]{DubinskiCarlberg1991, CroneEvrardRichstone1994,
  NavarroFrenkWhite1996, NavarroFrenkWhite1997, MooreEtal1999,
  BullockEtal2001a}. This `universal' halo density distribution is characterised
by a relatively shallow power-law behaviour in the inner parts, $\rho\sim
r^{-\gamma}$, with $\gamma\approx1$ typically inferred at the smallest resolved
radii, which steepens gradually to an extrapolated $\gamma\approx 3 - 4$ at
arbitrarily large radii.

A physical explanation, based on first principles, for the origin of such a
profile is still lacking, and there has been some considerable debate over the
exact functional form implied by the numerical studies. The fitting function
most commonly applied has the general form
\begin{equation}
  \label{eq:fitfunc}
  \rho(r) \propto
          r^{-\gamma_0}(r_s+r)^{\gamma_0-\gamma_\infty},
\end{equation}
where $\gamma_0$ and $\gamma_\infty$ are the power-law slopes of the central
cusp and in the limit $r\to\infty$, respectively, while $r_s$ is an appropriate
scale radius. \cite{DubinskiCarlberg1991} originally suggested the
\cite{Hernquist1990} profile, corresponding to $\gamma_0=1$ and
$\gamma_\infty=4$, while \cite*{NavarroFrenkWhite1996,NavarroFrenkWhite1997}
argued for $\gamma_0=1$ but $\gamma_\infty=3$ (the so-called `NFW' profile).
Subsequently, several studies have argued for a somewhat steeper central cusp
\citep[e.g.][]{MooreEtal1998, MooreEtal1999, GhignaEtal1998, GhignaEtal2000,
  FukushigeMakino1997, FukushigeMakino2001}, with one recent suite of
high-resolution simulations appearing to imply $\gamma_0\simeq1.16\pm0.14$
\citep{DiemandMooreStadel2004b}. However, even the highest-resolution simulation
to date \citep[][$\gamma_0=1.2$]{DiemandEtal2005} can resolve the halo structure
only to a fraction $10^{-3}$ of its virial radius, and this for a single halo
only. Much more common are numerical resolution limits several times larger than
that, leaving room for the possibility that halo densities might become \emph{
  shallower} than $r^{-1}$ at very small (`unobserved') radii
\citep{TaylorNavarro2001, PowerEtal2003, FukushigeKawaiMakino2004,
  HayashiEtal2004}, and perhaps even tend to a finite density at $r=0$ with no
cusp at all \citep{StoehrEtal2002, NavarroEtal2004, MerrittEtal2005}.
Alternatively, there might be no single, `universal' density slope in this limit
\citep{NavarroEtal2004, FukushigeKawaiMakino2004}. At the other extreme, there
are very few hard constraints on any limiting value of the density slope as
$r\to\infty$, as any halo is only well defined within a finite virial radius.

\begin{figure*}
  \centerline{\hfil
    \resizebox{82mm}{!}{\includegraphics{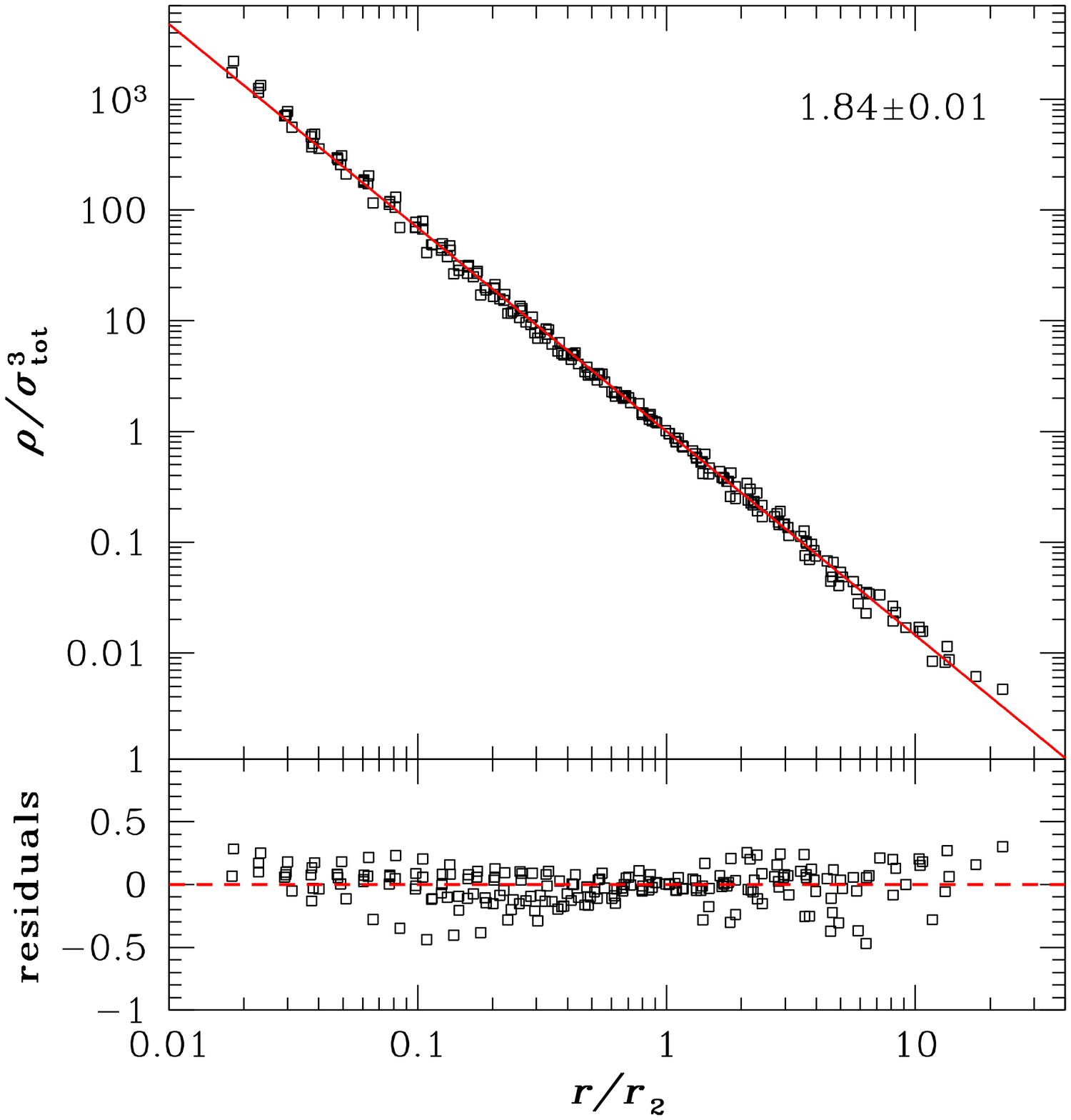}}\hspace{4mm}
    \resizebox{82mm}{!}{\includegraphics{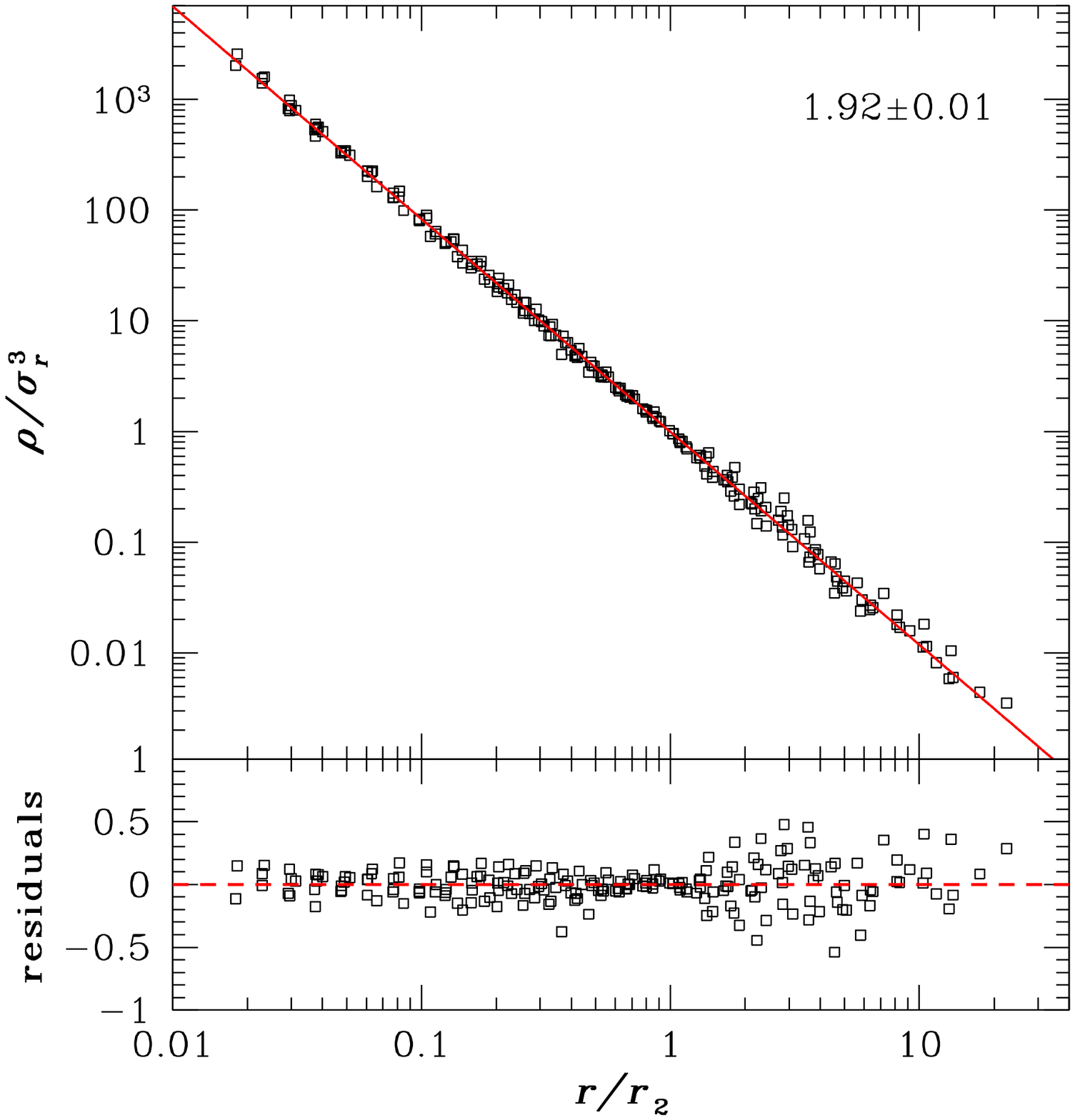}}
  }
  \caption{
    \label{fig:rhosig}
    Plots of $\rho/\sigma_{\mathrm{tot}}^3$ (\emph{left}) and $\rho/\sigma_r^3$
    (\emph{right}) as functions of radius for ten simulated CDM haloes from
    \citet{DiemandMooreStadel2004a, DiemandMooreStadel2004b}, namely the six
    cluster-sized haloes called A9, B9, C9, D12, E9, and F9 and the four
    galaxy-sized haloes G0, G1, G2, and G3. The radial coordinate in each halo
    has been scaled by the radius $r_2$ at which
    $\mathrm{d}\ln\rho/\mathrm{d}\ln r=-2$ for that halo, and both ratios
    $\rho/\sigma_{\mathrm{tot}}^3$ and $\rho/\sigma_r^3$ have been normalised by
    their respective values at $r_2$. The lines in the upper panels are
    power-law fits to the scaled data from all ten haloes combined.  The average
    negative slope and its uncertainty are indicated. In the bottom panels, we
    show the relative residuals $[(\rho/\sigma^3)
    -(\rho/\sigma^3)_{\mathrm{fit}}]/(\rho/\sigma^3)$.  The rms residual is
    essentially the same ($0.14$--$0.15$) whether $\sigma_{\mathrm{tot}}$ or
    $\sigma_r$ is involved, although in the latter case they are arguably
    distributed more uniformly about 0.  }
\end{figure*}

In this paper, we examine the question of dark-matter halo structure from a
slightly different viewpoint, opting to derive $\rho(r)$ from a simple dynamical
\emph{Ansatz}, rather than fitting a pre-set family of functions to simulated
density profiles. Our starting point is still empirical, however, being based on
another surprisingly uniform property of $N$-body haloes. As was first noted and
exploited by \cite{TaylorNavarro2001}, the ratio of the density $\rho(r)$ and
the cube of the velocity dispersion $\sigma(r)$ is a single power-law in radius
\begin{equation}
  \label{eq:rhosig}
   \rho(r)/\sigma(r)^3 \propto r^{-\alpha}
\end{equation}
over the full numerically resolved radial range. \citeauthor{TaylorNavarro2001}
originally found $\alpha=15/8=1.875$, which interestingly is the value predicted
by the classic similarity solution for spherical secondary infall
\citep{Bertschinger1985}. They then used this constraint to solve the Jeans
equation numerically (assuming an isotropic velocity distribution) for $\rho(r)$
and $\sigma(r)$ separately. Although their result for $\rho(r)$ is not of the
form (\ref{eq:fitfunc}) (it does tend to a shallow power law at the centre but
steepens rapidly outwards and falls to zero at a finite radius),
\citeauthor{TaylorNavarro2001} argued such a density distribution to be an
adequate description of simulation data inside the virial radius.
Unfortunately, there is no exact analytical expression for $\rho(r)$ in the case
$\alpha=1.875$, and this approach has not been used for fitting any data. Other
studies have subsequently confirmed that $\rho/\sigma^3$ is a power law in
radius, but estimates of the exponent differ somewhat from the Taylor-Navarro
value: $\alpha=1.95$ or $1.90\pm0.05$ according to
\cite*{RasiaTormenMoscardini2004} and \cite{AscasibarEtal2004}, respectively.

Our main aim is to investigate in more detail the structure and dynamics of
spherical dark-matter haloes that follow a `$\rho$-$\sigma$ relation' of the
basic type given in equation (\ref{eq:rhosig}), allowing at least initially for
arbitrary values of $\alpha$.  Thus, in Figure~\ref{fig:rhosig} we show this
relation as defined by ten CDM haloes simulated by
\cite*{DiemandMooreStadel2004a, DiemandMooreStadel2004b}, the details of which
were kindly provided to us by J\"urg~Diemand. These include four galaxy-sized
haloes and six cluster-sized haloes, with virial masses ranging from
$10^{12}\,M_\odot$ to $10^{15}\,M_\odot$. All are dynamically relaxed. The left
panels of Fig.~\ref{fig:rhosig} examine the ratio $\rho/\sigma_{\mathrm{tot}}^3$
as a function of radius in these haloes, where the total one-dimensional
velocity dispersion is $\sigma_{\mathrm{tot}}^2 \equiv
\frac{1}{3}(\sigma_r^2+\sigma_\theta^2+\sigma_\phi^2)$.  The right panels look
at the quantity $\rho/\sigma_r^3$ vs.~$r$, where $\sigma_r$ is the velocity
dispersion in the radial direction only. Power-law fits to each of these
profiles are drawn, and the residuals from the fits are shown in the bottom
panels. These demonstrate that the ratio $\rho/\sigma_r^3$ follows a power law
in radius \emph{at least} as closely as $\rho/\sigma_{\mathrm{tot}}^3$ does,
although the fitted slopes differ slightly between the two cases. Fits for each
of the ten haloes individually yield power-law slopes that can differ by
$\pm4\%$ from the average values in Fig.~\ref{fig:rhosig}.  Whether this scatter
is real or simply a reflection of numerical uncertainties is unclear, but it is
certainly rather modest.

To the extent that either of the $\rho$-$\sigma$ relations illustrated in
Fig.~\ref{fig:rhosig} is `universal', and insofar as dark-matter haloes are in
equilibrium, imposing a dynamical constraint along the lines of equation
(\ref{eq:rhosig}) to solve the spherical Jeans equation, as
\citeauthor{TaylorNavarro2001} originally did, should lead directly to a
`universal' density profile. While we have no physical argument for the
fundamental origin of the precise power-law behaviour in Figure \ref{fig:rhosig}
(though clearly it must be related to the initial conditions and the formation
via violent relaxation), it is much simpler to characterise than the density
profile itself.  Moreover, a Jeans-equation approach allows explicitly for a
simultaneous exploration of velocity anisotropy inside haloes---an issue which
to date has been largely divorced from empirical descriptions of the halo
density profiles.

It is well known that the velocity distributions in dark-matter haloes are not
isotropic. We characterise velocity anisotropy using Binney's parameter
\begin{equation}
  \label{eq:beta}
  \beta(r) = 1 - \frac{\sigma_\theta^2+\sigma_\phi^2}{2\sigma_r^2},
\end{equation}
such that $0<\beta\le1$ corresponds to radial anisotropy and $\beta<0$ signifies
a tangentially biased velocity distribution. It is typically found that
$\beta\approx0$ (isotropy) at the centres of haloes and gradually increases
outwards (reflecting radial anisotropy), reaching levels of $\beta\approx0.5$
around the virial radius \citep[e.g.,][]{ColinKlypinKravtsov2000,
  FukushigeMakino2001}. Indeed, it has been suggested \citep{ColeLacey1996,
  CarlbergEtal1997} that there exists a `universal' anisotropy profile in
dark-matter haloes. Very closely related to such an idea is the recent claim by
\cite{HansenMoore2005}, that $\beta(r)$ depends roughly linearly on the
logarithmic density gradient
\begin{equation}
  \label{eq:gamma}
  \gamma(r) \equiv - \mathrm{d}\ln\rho/ \mathrm{d}\ln r
\end{equation}
with essentially the same, constant slope $\mathrm{d}\beta/\mathrm{d}\gamma$
holding for a variety of end-products of violent relaxation processes (merger
remnants, dark-matter haloes, collapse remnants).

With these points particularly in mind, we base our analysis on the assumption
that halo density and the \emph{radial} component of velocity dispersion
$\sigma_r$ (rather than $\sigma_{\mathrm{tot}}$) are connected through a
power-law relation of the general form
\begin{equation}
  \label{eq:alpha}
  \frac{\rho}{\sigma_r^\epsilon}(r) =
     \frac{\rho_0}{\sigma_{r,0}^\epsilon}\,
        \left(\frac{r}{r_0}\right)^{-\alpha},
\end{equation}
where $r_0$ is any convenient reference radius, $\rho_0=\rho(r_0)$, and
$\sigma_{r,0} = \sigma_r(r_0)$.  Intuitively, as well as on the basis of Figure
\ref{fig:rhosig}, it seems most natural to expect the exponent $\epsilon$ in
this equation to be $\epsilon=3$; but it adds little complication to allow for
the possibility \citep{Hansen2004} that a slightly different value might provide
a still more accurate description of simulated haloes. The choice of $\sigma_r$
as the velocity dispersion to work with is arguably more natural than
$\sigma_{\mathrm{tot}}$, since $\beta(r)$ and $\sigma_r$ appear separately in
the spherical Jeans equation but $\sigma_{\mathrm{tot}}^2 = \sigma_r^2
(1-2\beta/3)$. Technically, this choice ultimately allows for a more tractable
inclusion of velocity anisotropy; empirically, it is obviously well justified by
Figure \ref{fig:rhosig}.

We begin in Section~\ref{sec:isotropic} with an investigation into which
spherical density profiles satisfy the Jeans equation and obey the
\emph{Ansatz}~(\ref{eq:alpha}) under the restrictions of velocity isotropy
($\beta\equiv0$ and $\sigma_{\mathrm{tot}}\equiv\sigma_r$) and a fixed
$\epsilon=3$. The problem is then identical to the one first considered by
\cite{TaylorNavarro2001}, and our approach is rooted in theirs, but we also draw
on some aspects of the considerations by \cite{WilliamsEtal2004}. However,
unlike these authors, we explore the full solution space of the problem. We find
that only very few of the many possible solutions correspond to realistic
density models for simulated dark-matter haloes, and in fact only one solution,
which occurs for a `critical' value of $\alpha$, is of practical importance.

We then proceed in Section~\ref{sec:aniso} to consider the more realistic case
of anisotropic velocity distributions and allow for general values of $\epsilon$
in equation (\ref{eq:alpha}). We show that in the case of an anisotropy
parameter $\beta$ that depends linearly on $\gamma$ (including constant
anisotropy as a special case), and for any $\epsilon$, the solutions of the
Jeans equation under our adopted constraint are exact analogues of those in the
$\beta\equiv0$, $\epsilon=3$ case. In particular, for each pair ($\epsilon$,
$\beta(r=0)$) only one physical solution of practical relevance exists, which
again occurs at a `critical' $\alpha$ value. These solutions have fully
analytical density, mass, and velocity-dispersion profiles with power-law
asymptotes at small and large radii.

In Section~\ref{sec:simhalo} we compare our analytical profiles to the
`observed' density, velocity dispersion, and anisotropy profiles in haloes
simulated by \cite{DiemandMooreStadel2004a, DiemandMooreStadel2004b,
  DiemandEtal2005}, finding good agreement in general. Finally,
Section~\ref{sec:summary} discusses our findings and summarises the paper.

\section{The isotropic case}
\label{sec:isotropic}
Our underlying assumption is that some version of the general $\rho$-$\sigma$
relation in equation (\ref{eq:alpha}) holds for dark-matter haloes. Before
allowing for this level of generality, however, there is much insight to be
gained from beginning with a more specialised case, in which the velocity
distribution is isotropic and $\epsilon=3$. Then, as in
\cite{TaylorNavarro2001}, \cite{WilliamsEtal2004}, and \cite{Hansen2004}, we
have
\begin{equation}
  \label{eq:special}
  \frac{\rho}{\sigma_r^3}
  = \frac{\rho_0}{\sigma_{r,0}^3}\,\left(\frac{r}{r_0}\right)^{-\alpha}.
\end{equation}
The Jeans equation for a spherical, self-gravitating collisionless system with
isotropic velocity distribution is
\begin{equation}
  \label{eq:jeans:a}
  \frac{\mathrm{d}(\rho\sigma_r^2)}{\mathrm{d}r}
  = - \rho(r)\,\frac{GM(r)}{r^2}
\end{equation}
with $M(r)=4 \pi \int_0^r u^2 \rho(u)\,\mathrm{d}u$.  Following
\citeauthor{TaylorNavarro2001}, we solve equation (\ref{eq:special}) for
$\sigma_r$, insert it in equation~(\ref{eq:jeans:a}), and differentiate again to
obtain
\begin{equation}
  \label{eq:iso:a}
  \frac{1}{x^2 y}  
  \frac{\mathrm{d}}{\mathrm{d}x} \left[
    \frac{x^2}{y} 
    \frac{\mathrm{d}}{\mathrm{d}x}(y^{5/3} x^{2\alpha/3})\right] = -\kappa.
\end{equation}
Here, $x\equiv r/r_0$ and $y\equiv\rho/\rho_0$ are dimensionless variables, and
\begin{equation}
  \label{eq:kappa}
  \kappa \equiv 4\pi G \rho_0 r_0^2/\sigma_{r,0}^2
\end{equation}
is a dimensionless measure of the velocity dispersion scale.
\citeauthor{TaylorNavarro2001} studied the solutions of
equation~(\ref{eq:iso:a}) for the particular value $\alpha=15/8$ by numerical
integration. It proves useful, however, to first re-write the problem in terms
of the (negative) logarithmic density slope $\gamma(r)$, as defined in
(\ref{eq:gamma}). Equation (\ref{eq:iso:a}) then reads
\begin{equation}
  \label{eq:iso:b}
  \gamma^\prime - \tfrac{2}{3}
  (\gamma-\gamma_a)\,
  (\gamma-\gamma_b)
  = \tfrac{3}{5}\kappa x^{2-2\alpha/3} y^{1/3},
\end{equation}
where a prime denotes differentiation with respect to $\ln x$ and
\begin{equation}
  \label{eq:gggab}
  \gamma_a = \tfrac{2}{5}\,\alpha, \qquad
  \gamma_b = \alpha + \tfrac{3}{2}.
\end{equation}

The objective of this Section is to investigate the solution space of
equation~(\ref{eq:iso:b}). First, note that its r.h.s.\ becomes constant for
$y=x^{-\gamma_1}$, with
\begin{equation}
  \label{eq:ggg1}
  \gamma_1 = 6 - 2 \alpha.
\end{equation}
In fact, as was already noted by \citeauthor{TaylorNavarro2001}, this singular
density profile is a solution to equation~(\ref{eq:iso:b}) and, for $\alpha=2$
corresponds to the well-known singular isothermal sphere. For $3/2<\alpha<5/2$
(which covers our regime of interest; see Figure \ref{fig:rhosig} above),
\begin{equation}
  \gamma_a < \gamma_1 < \gamma_b 
\end{equation}
and thus, following \citeauthor{TaylorNavarro2001}, we choose to identify the
reference radius $r_0$ in equation~(\ref{eq:special}) as that at which the
(negative) density slope equals $\gamma_1$, i.e., $\gamma_1=\gamma(x=1)$.  In
this way, $r_0$ is well-defined for all realistic solutions\footnote{As we shall
  see, for some solutions $\gamma(r)=\gamma_1$ at more than just one radius. In
  this case, we pick that radius which in addition maximises $\gamma^\prime$.}.
Moreover, it is then obvious from this equation that the constant $\kappa$ is
effectively a measure of $\gamma^\prime$ at $x=1$.

\begin{figure*}
  \centerline{\hfil
    \resizebox{64mm}{!}{\includegraphics{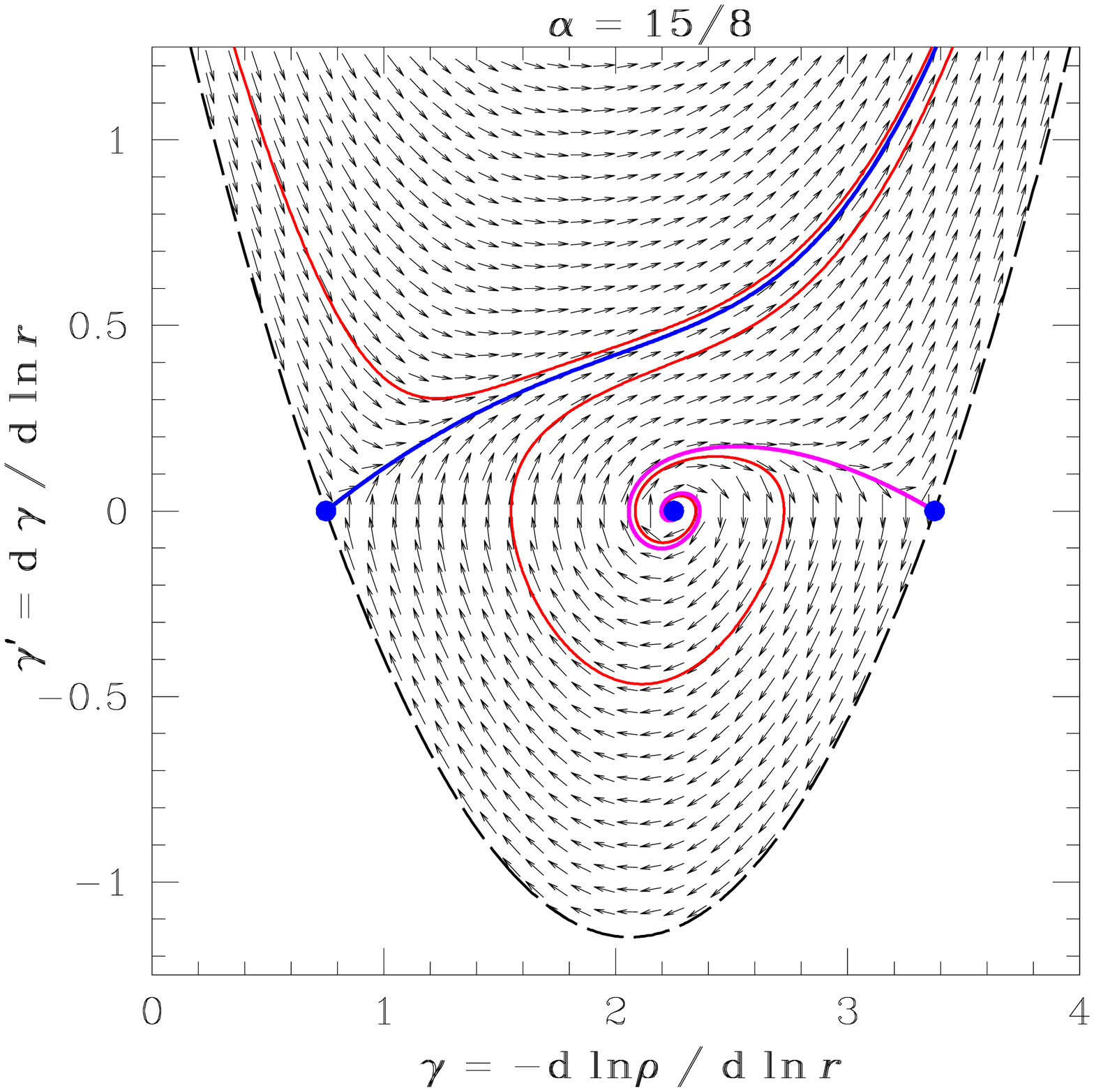}}\hspace*{-8mm}
    \resizebox{64mm}{!}{\includegraphics{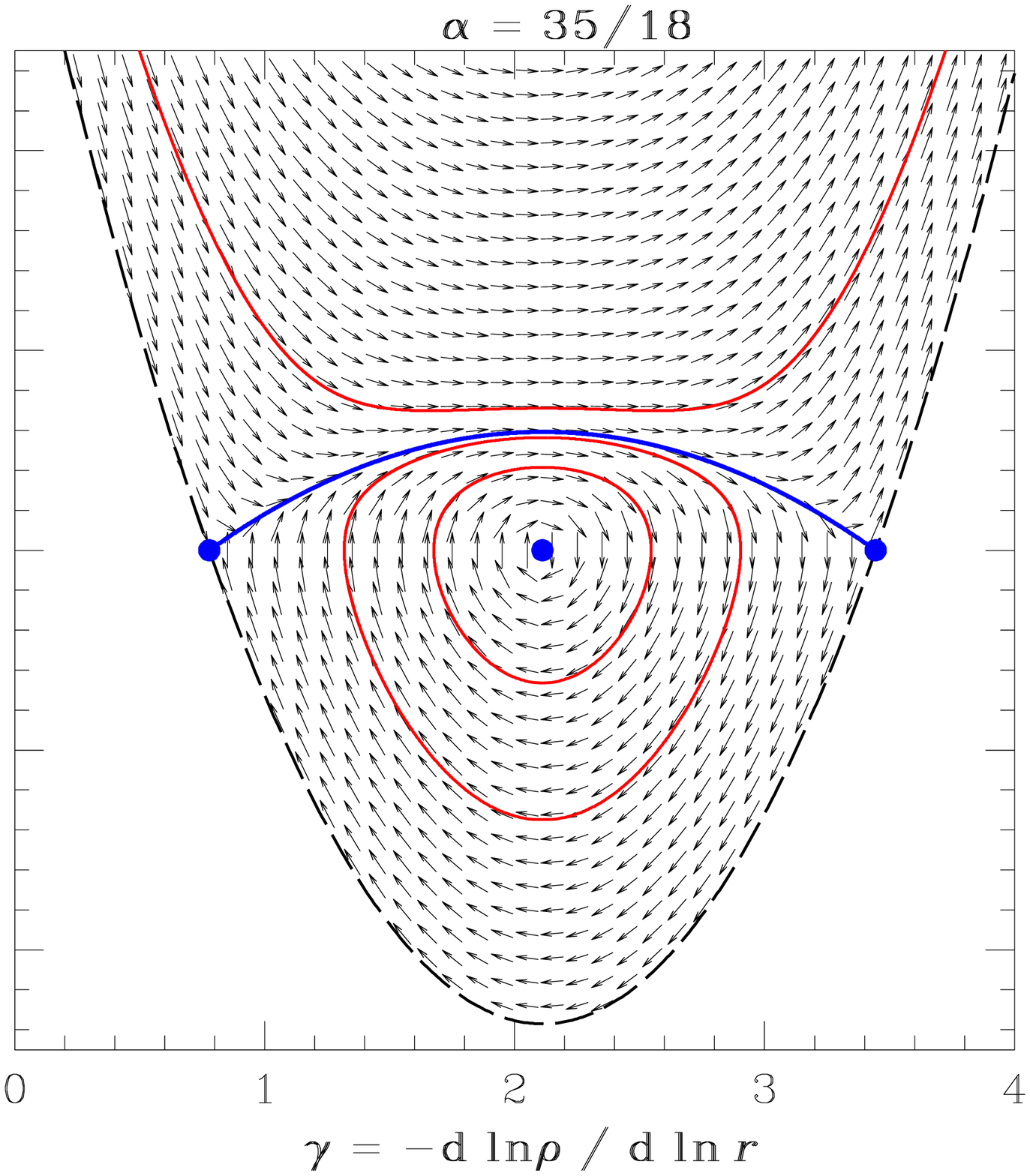}}\hspace*{-8mm}
    \resizebox{64mm}{!}{\includegraphics{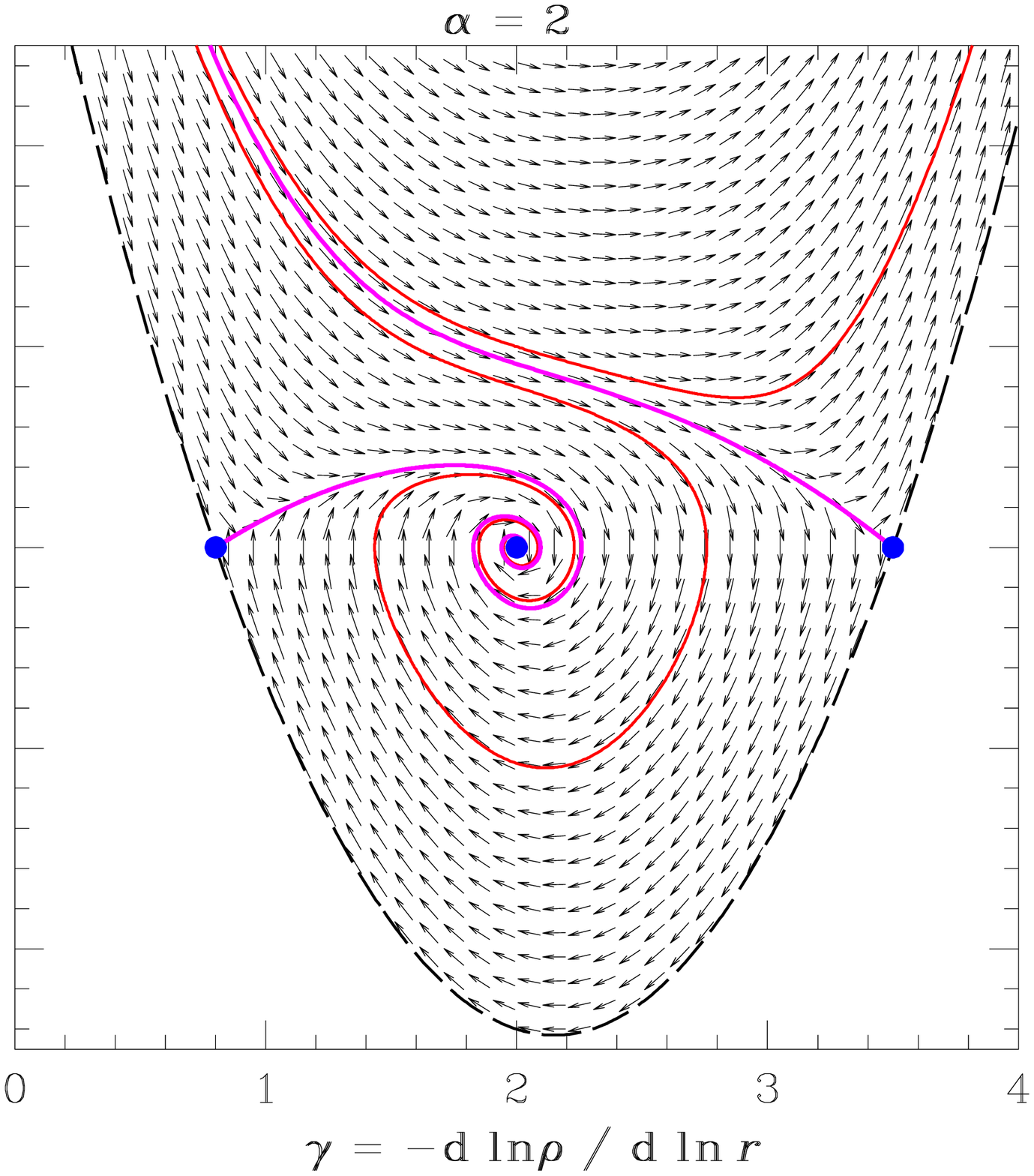}}\hfil
  }
  \caption{
    \label{fig:flow}
    Flow diagrams of equation~(\ref{eq:iso:c}), describing isotropic density
    profiles with $\rho/\sigma_r^3\propto r^{-\alpha}$, in the $(\gamma,
    \gamma^\prime)$ phase space (the length of the arrows has been normalised
    for ease of reading) for three different values of $\alpha$. The flow
    vanishes at the three fixed points (\emph{dots}), where $\gamma^\prime=0$
    and $\gamma$ is a root of the right-hand side of equation~(\ref{eq:iso:c}).
    Some solutions to equation~(\ref{eq:iso:c}) are given as curves. Physical
    solutions to the original equation~(\ref{eq:iso:b}) must lie above the
    (\emph{dashed}) parabola, which is defined in equation~(\ref{eq:gamma:min}).
    Solutions with $\gamma^\prime<0$ or $\gamma<0$ at any point (\emph{red} and
    \emph{magenta} curves) are unrealistic.  There is exactly one realistic and
    physical solution for every $\alpha\le35/18$ (\emph{blue} curve) and none
    for larger $\alpha$.  At small radii this solution's density profile
    approaches the power-law $\rho\propto r^{-2\alpha/5}$, while at large radii
    the density is truncated at a finite radius for $\alpha<35/18$, and
    approaches $\rho\propto r^{-\alpha-3/2}=r^{-31/9}$ for $\alpha=35/18$.  }
\end{figure*}

Let us now consider equation~(\ref{eq:iso:b}) in some detail, in particular the
possible behaviour of any solutions at very small and large radii (see also
\citeauthor{WilliamsEtal2004}). It is straightforward to verify that in the
limit $r\to0$ only three asymptotes are possible: either $\gamma\to\gamma_a$
(both sides of the equation vanish, so long as $\alpha<5/2$); or
$\gamma\to\gamma_1$ (both sides approach a constant); or the density has a
central hole with $y\equiv0$ inside a non-zero radius (note that
$\gamma\to\gamma_b$ is not possible, since the r.h.s.\ would then diverge for
$\alpha>3/2$).\footnote{\cite{Hansen2004}, in his analysis of the same problem,
  mentions the asymptote $\gamma\to\gamma_1$ in the limit $r\to 0$, but misses
  the important possibility $\gamma\to\gamma_a$.} Of these three, the latter is
unrealistic, whereas $\gamma\to\gamma_1$ is inconsistent with simulated haloes,
which have shallower inner density slopes.  This leaves $\gamma\to\gamma_a$ as
the only interesting option. We thus expect that, for any given value of
$\alpha$, there exists one solution with $\gamma\to\gamma_a$ as $r\to0$. The
solution with this limiting behaviour is associated with a unique value of
$\kappa$, which depends on $\alpha$ through equation (\ref{eq:iso:b}).  In fact,
this very case corresponds to the `critical $\kappa$' solution developed by
\citeauthor{TaylorNavarro2001} for their specified $\alpha=15/8$.

In the limit $r\to\infty$, also only three possible asymptotes exist: either
$\gamma\to\gamma_b$ (both sides of the equation vanish, if $\alpha>3/2$), or
$\gamma\to\gamma_1$ (both sides approach a constant again), or the density has
an outer truncation with $y\equiv0$ beyond a finite radius (note that
$\gamma\to\gamma_a$ is not possible, as the r.h.s.\ then diverges if
$\alpha<5/2$). All of these three are physically meaningful (apart from the fact
that for $\gamma\to\gamma_1$ the mass formally diverges, similarly to the
isothermal sphere). \citeauthor{TaylorNavarro2001}'s `critical $\kappa$'
solution for $\alpha=15/8$ is one with an outer truncation.

In order to gain more insight into the solution space of
equation~(\ref{eq:iso:b}), we now follow \citeauthor{WilliamsEtal2004} and
differentiate it yet again (with respect to $\ln x$) to obtain
\begin{equation}
  \label{eq:iso:c}
  \gamma^{\prime\prime}-\gamma^\prime\,
  \Big(\gamma - \tfrac{1}{3}[2\gamma_a+2\gamma_b-\gamma_1]\Big) 
  = \tfrac{2}{9} (\gamma-\gamma_a)\, (\gamma-\gamma_b)\, (\gamma-\gamma_1),
\end{equation}
equivalent to their equation~(2.2). In Figure~\ref{fig:flow}, we plot flow
diagrams of the $(\gamma,\gamma^\prime)$ phase space for three representative
values of $\alpha$. The \emph{dashed} parabola in each panel corresponds to
\begin{equation}
  \label{eq:gamma:min}
  \gamma^\prime = \gamma^\prime_{\mathrm{min}} \equiv 
  \tfrac{2}{3} (\gamma-\gamma_a)\, (\gamma-\gamma_b).
\end{equation}
While this actually is a solution to (\ref{eq:iso:c}), as already discovered by
\citeauthor{WilliamsEtal2004}, it is not a valid solution to the original,
physical problem. This is because the constant $\kappa > 0$ by definition, so
equation (\ref{eq:iso:b}) requires that $\gamma^\prime \ge
\gamma^\prime_{\mathrm{min}}$ in order for its r.h.s.\ to be non-negative.  In
fact, the equality $\gamma^\prime=\gamma^\prime_{\mathrm{min}}$ is only allowed
in either of the limits $\gamma\to\gamma_a$ as $r\to 0$ or $\gamma\to\gamma_b$
as $r\to\infty$ (see the discussion above). Similarly, any solutions below the
parabola~(\ref{eq:gamma:min}) require $\kappa<0$ and are unphysical.  By
contrast, all solutions above the parabola~(\ref{eq:gamma:min}) correspond to
potentially viable solutions of the original equation (\ref{eq:iso:b}).

In all panels of Figure \ref{fig:flow}, the flow vanishes at the three fixed
points $(\gamma,\gamma^\prime)=(\gamma_a,0)$, $(\gamma_1,0)$, and
$(\gamma_b,0)$, which are plotted as \emph{blue} dots. The central fixed point
is stable and corresponds to the singular solution $y=x^{-\gamma_1}$ discussed
above. Solutions with $\gamma^{\prime\prime}>0$ (arrows pointing upwards) at any
$\gamma>\gamma_1$ approach, in the limit $\gamma\to\infty$, the upper right
branch of the parabola $\gamma^\prime=\gamma^\prime_{\mathrm{min}}$. This
corresponds to the case, discussed above, of an outer truncation to the density.
More precisely, integrating equation (\ref{eq:gamma:min}) twice yields the
asymptotic density profile
\begin{equation}
  \rho \to
      r^{-\gamma_b} \big(r_t^{2(\gamma_b-\gamma_a)/3} -
          r^{2(\gamma_b-\gamma_a)/3}\big)^{3/2}
\end{equation}
as $r$ approaches some finite radius $r_t$. For $r \ge r_t$, $\rho\equiv0$.
Conversely, solutions with $\gamma^{\prime\prime}<0$ (arrows pointing downwards)
at any $\gamma<\gamma_1$ approach, in the limit $\gamma\to-\infty$, the upper
left branch of the parabola. This corresponds to the case, mentioned above, of
an inner truncation, with the asymptotic density profile
\begin{equation}
  \rho \to r^{-\gamma_a} 
  \big(r_h^{2(\gamma_a-\gamma_b)/3} - r^{2(\gamma_a-\gamma_b)/3}\big)^{3/2}
\end{equation}
as $r$ approaches some radius $r_h>0$. For $r\le r_h$, $\rho\equiv 0$. Thus
solutions for which $\gamma\to\infty$ at large radii have a finite outer
truncation radius, and solutions for which $\gamma\to-\infty$ at small radii
have a finite-sized inner hole. These latter solutions are clearly unrealistic
and also unphysical (the isotropic distribution function must become negative to
account for the hole). Furthermore, solutions that ever visit $\gamma^\prime<0$
are hardly realistic (their density profiles become \emph{shallower} towards
larger radii, at least over some radial range).

Apart from this---as \cite{WilliamsEtal2004} have also noted---the generic
behaviour of the solutions to equation (\ref{eq:iso:c}) depends on whether
$\alpha$ is greater than, less than, or equal to a critical value
$\alpha_{\mathrm{crit}}$ for which the inner and outer fixed points $\gamma_a$
and $\gamma_b$ are equidistant from the central $\gamma_1$.  Referring to
equations (\ref{eq:gggab}) and (\ref{eq:ggg1}),
$\gamma_1=\tfrac{1}{2}(\gamma_a+\gamma_b)$ requires $\alpha_{\mathrm{crit}} =
35/18 = 1.9\overline{4}$.

The behaviour of the solutions for $\alpha>\alpha_{\mathrm{crit}}$ is
exemplified in the right panel of Fig.~\ref{fig:flow}. The solutions are split
into two families by the one (\emph{upper magenta}) which ends at the fixed
point at $\gamma=\gamma_b$, so that $\rho\propto r^{-\gamma_b}$ as $r\to\infty$.
Solutions lying above this curve correspond to density profiles with a central
hole and an outer truncation; those below usually also have a central hole but
at larger radii perform damped `oscillations' about $\gamma=\gamma_1$,
eventually approaching $\rho\propto r^{-\gamma_1}$ as $r\to\infty$. All of these
solutions are unrealistic, since they possess inner density holes. The only
exception is a limiting solution (\emph{lower magenta}) which starts from the
fixed point at $\gamma=\gamma_a$ (i.e., $\rho\propto r^{-\gamma_a}$ as $r\to0$)
and slowly approaches $\gamma=\gamma_1$ as $r\to\infty$. However, this solution
is still not a viable description of dark-matter haloes, since its density
profile is too shallow at large radii.

The behaviour of the solutions when $\alpha<\alpha_{\mathrm{crit}}$ is
illustrated in the left panel of Fig.~\ref{fig:flow} for $\alpha=15/8$, the case
previously studied by \citeauthor{TaylorNavarro2001}. The situation is in some
sense a mirrored version of that for $\alpha>\alpha_{\mathrm{crit}}$. The
solution space is again divided into two families, now by the solution
(\emph{blue}) starting from the left fixed point at $\gamma=\gamma_a$ at $r=0$.
Solutions above this one again correspond to density profiles with a central
hole and an outer truncation; those below it start from an `oscillation' about
the power-law $\rho\propto r^{-\gamma_1}$ in the limit $r\to0$, eventually
steepening outwards and generally being truncated at a finite large radius. The
limit of this family is the solution (\emph{magenta}) which instead of an outer
truncation has a power-law fall-off with $\rho\propto r^{-\gamma_b}$ as
$r\to\infty$.

For any $\alpha<\alpha_{\mathrm{crit}}$, the (\emph{blue}) solution separating
the two families just described \emph{is} a potentially realistic model for
dark-matter haloes, starting as it does from a shallow power-law cusp
$\rho\propto r^{-\gamma_a}$ at $r\to0$ (with $\gamma_a=2\alpha/5<7/9$ for
$\alpha<35/18$) and steepening outwards to reach $\rho=0$ at a finite radius.
This is the model identified by \citeauthor{TaylorNavarro2001} as their
`critical $\kappa$' solution in the specific case $\alpha=15/8$. Given this
$\alpha$, numerical integration of equation (\ref{eq:iso:c}) outwards from
$\gamma=\gamma_a=3/4$ at $x=0$ yields $\kappa=2.674$, to be compared with the
value $\kappa=2.678$ found by \citeauthor{TaylorNavarro2001} through
trial-and-error integration of equation (\ref{eq:iso:a}) starting from $x=1$.

The middle panel of Figure \ref{fig:flow} depicts the special situation
$\alpha=\alpha_{\mathrm{crit}}$, for which the flow in $(\gamma,\gamma^\prime)$
phase-space is symmetric with respect to the transformation $r\to r^{-1}$ and
$\gamma\to2\gamma_1-\gamma$ (because $\gamma_a+\gamma_b=2\gamma_1$ by definition
of $\alpha_{\mathrm{crit}}$) In this case, there exists a first integral,
\begin{equation}
  \label{eq:iso:K}
  K = 
  \Big(\gamma^\prime+\tfrac{1}{6}(\gamma-\gamma_a)(\gamma-\gamma_b)\Big)\,
  \Big(\gamma^\prime-\tfrac{2}{3}(\gamma-\gamma_a)(\gamma-\gamma_b)\Big)^4\ ,
\end{equation}
which is conserved by any solution of equation (\ref{eq:iso:c}) with
$\alpha=\alpha_{\mathrm{crit}}$. In particular, $K=0$ for the solution
$\gamma^\prime=\gamma^\prime_{\mathrm{min}}(\gamma)$ (see
eq.~[\ref{eq:gamma:min}]) and also for
\begin{equation}
  \label{eq:iso:solution}
  \gamma^\prime = -\tfrac{1}{6}(\gamma-\gamma_a)(\gamma-\gamma_b)
\end{equation}
which is plotted as the \emph{blue} curve in the middle panel of Figure
\ref{fig:flow}. It again divides the solution space into two, although now
solutions above the curve (\ref{eq:iso:solution}) always have both an inner hole
and an outer truncation to the density profile, while solutions below the curve
undergo undamped `oscillations' about $\rho\propto r^{-\gamma_1}$, never
settling to an asymptote in either limit $r\to0$ or $r\to\infty$.

The solution (\ref{eq:iso:solution}) for $\alpha=\alpha_{\mathrm{crit}}=35/18$
is particularly appealing since it starts from a shallow power law
$\gamma\to\gamma_a=7/9$ in the limit $r\to0$ and tends to the steeper
$\gamma\to\gamma_b=31/9$ as $r\to\infty$. The combination of these features is
reminiscent of dark-matter haloes. This single solution with
$\alpha=\alpha_{\mathrm{crit}}$ is the \emph{only} one, of all the solutions for
any $\alpha$, whose density has both inner and outer power-law asymptotes and
monotonically increasing $\gamma$. It is also the only one we have found which
is simple enough that almost all its physical properties can be developed
analytically: equation (\ref{eq:iso:solution}) is easily integrated to give
$\gamma(x)$ and subsequently $y(x)\propto \rho(r)$ as simple functions, which
then allows us to evaluate $\sigma_r(r)$, the enclosed mass profile $M(r)$, and
the circular velocity $V_c(r)$.

To aid in obtaining these basic results, we first substitute equation
(\ref{eq:iso:solution}) into equation (\ref{eq:iso:b}) and evaluate the result
at $x=y=1$ (where $\gamma=\gamma_1$ by definition) to find
\begin{subequations} 
  \label{eq:iso:model}
  \begin{equation}\label{eq:kappa:crit}
    \kappa=\tfrac{25}{18}(\gamma_1-\gamma_a)(\gamma_b-\gamma_1)
    = \tfrac{200}{81}
  \end{equation}
  (for $\alpha=\alpha_{\mathrm{crit}}$).  With this in hand, we obtain
  \begin{eqnarray} 
  \label{eq:iso:density}
  \rho(r) &=& \frac{5}{9}\frac{M_{\mathrm{tot}}}{\pi r_0^3}
  x^{-7/9} \Big(1 + x^{4/9}\Big)^{-6},
  \\
  \label{eq:iso:gamma}
  \gamma(r) &=& \frac{\tfrac{7}{9} + \tfrac{31}{9} x^{4/9}}{1 + x^{4/9}},
  \\
  \label{eq:iso:sigma}
  \sigma_r^2(r) &=& \frac{9}{40}\frac{GM_{\mathrm{tot}}}{r_0}
  x^{-1} \left(\frac{x^{4/9}}{1+x^{4/9}} \right)^{4},
  \\
  \label{eq:iso:mass}
  M(r) &=& M_{\mathrm{tot}} \left(\frac{x^{4/9}}{1+x^{4/9}} \right)^{5},
  \\
  \label{eq:iso:vcirc}
  V_c^2(r) & = & \frac{GM_{\mathrm{tot}}}{r_0} x^{-1}
  \left(\frac{x^{4/9}}{1+x^{4/9}} \right)^{5},
\end{eqnarray}
where $x\equiv r/r_0$ as usual. We have replaced $\rho_0$ with the total mass
$M_{\mathrm{tot}}=(576\pi/5)\,\rho_0r_0^3$; and the value of $\kappa$ in
equation (\ref{eq:kappa:crit}) has been used with the basic definition
(\ref{eq:kappa}) to eliminate the normalisation $\sigma_{r,0}$ from equation
(\ref{eq:iso:sigma}). Note that, because this solution has a finite total mass,
both $\sigma_r^2(r)$ and $V_c^2(r)$ fall off as $r^{-1}$ in the limit
$r\to\infty$. At the same time, both quantities vanish at $r=0$, and thus each
profile peaks at a finite radius. This happens at $x=(7/9)^{9/4}$ for
$\sigma_r^2(r)$, and at $x=(11/9)^{9/4}$ for $V_c^2(r)$. Finally, the
gravitational potential follows from integrating
$\mathrm{d}\Phi/\mathrm{d}r=GM(r)/r^2$:
\begin{equation}
  \label{eq:iso:pot}
  \Phi(r)= - \frac{9}{4} \frac{G\,M_{\mathrm{tot}}}{r_0}
             B_{\frac{1}{1+x^{4/9}}}\left(\tfrac{9}{4},\tfrac{11}{4}\right)\ ,
\end{equation}
\end{subequations} 
where $B_u(p,q)\equiv\int_0^u t^{p-1}(1-t)^{q-1}\mathrm{d}t$ is the incomplete
beta function \citep[e.g.][\S6.4]{PressEtal1992}. For an alternative form and
asymptotic limits of $\Phi$, see equation~(\ref{eq:aniso:pot:alt}) and the
following text.

Apart from its convenient and unique analytical properties, this solution of the
Jeans equation is additionally of interest because it corresponds to a
$\rho$-$\sigma$ relation of the type in equation (\ref{eq:special}) with
$\alpha=\alpha_{\mathrm{crit}}=1.9\overline{4}$, remarkably close to the
exponent actually found for the simulated dark-matter haloes shown in
Fig.~\ref{fig:rhosig} above. An obvious caveat is that our development in this
Section has assumed an isotropic velocity distribution, which is known to be
incorrect. Nevertheless, having characterised the solution space of equation
(\ref{eq:iso:a}) or (\ref{eq:iso:b}) for this restricted case, it turns out to
be straightforward to allow for realistic velocity anisotropies and at the same
time investigate other values of $\epsilon$ in the full $\rho$-$\sigma$ relation
of equation (\ref{eq:alpha}). As we will now show, the specialised isotropic
solution of equations (\ref{eq:iso:model}) is in fact just one of a larger, more
general family of analytical solutions to the Jeans equation.

\section{The anisotropic case}
\label{sec:aniso}
We now return to the more general form of our basic assumption~(\ref{eq:alpha}),
namely
\[
  \frac{\rho}{\sigma_r^\epsilon}(r) =
     \frac{\rho_0}{\sigma_{r,0}^\epsilon}\,
        \left(\frac{r}{r_0}\right)^{-\alpha},
\]
and seek to solve the spherical Jeans equation in the general form
\begin{equation}
  \label{eq:jeans:b}
  \frac{\mathrm{d}}{\mathrm{d}r}\,\rho\sigma_r^2 +
    \frac{2\beta(r)}{r}\,\rho\sigma_r^2
    = - \rho(r)\,\frac{GM(r)}{r^2}
\end{equation}
with anisotropy parameter $\beta$ as defined in equation~(\ref{eq:beta}).
By the same method as in the last Section, we then find the generalisation of
equation~(\ref{eq:iso:b}) to be
\begin{eqnarray}
  \left(\gamma^\prime -\frac{2\epsilon}{2+\epsilon}\,\beta^\prime\right) &-&
  \frac{2}{\epsilon}\left(\gamma-\alpha-\frac{\epsilon}{2}\right)
  \left(
  \gamma-\frac{2\epsilon}{2+\epsilon}\,\beta-\frac{2\alpha}{2+\epsilon}
  \right) \nonumber \\
  \label{eq:aniso:a}
  &=& \frac{\epsilon}{2+\epsilon}\,\kappa\, x^{2-2\alpha/\epsilon}
  y^{1-2/\epsilon}.
\end{eqnarray}
Differentiating again to obtain the equivalent of equation (\ref{eq:iso:c})
leads to a number of nonlinear terms involving up to the second derivative
$\beta^{\prime\prime}$, including cross-terms of the type $\gamma\beta^\prime$
and $\beta\gamma^\prime$.  These terms cancel exactly, however, if and only if
$\beta$ depends linearly on $\gamma$. That is, the structure of the Jeans
equation itself naturally suggests that we stipulate the relationship
\begin{equation}
  \label{eq:beta:linear}
  \beta=\beta_0+b(\gamma-\gamma_a)\ ,
\end{equation}
for $b$ and $\gamma_a$ constants. The definition of $\gamma_a$ follows
from substituting the expression (\ref{eq:beta:linear}) into
equation (\ref{eq:aniso:a}), yielding
\begin{equation}
  \label{eq:aniso:b}
  \gamma^\prime - \frac{2}{\epsilon}\,
  (\gamma-\gamma_a)\,
  (\gamma-\gamma_b)
  = \frac{\epsilon}{2+\epsilon-2\epsilon b}\,\kappa\, x^{2-2\alpha/\epsilon}
  y^{1-2/\epsilon},
\end{equation}
if
\begin{equation}
  \label{eq:aniso:gamma:a}
  \gamma_a = \frac{2\alpha}{2+\epsilon} +
             \frac{2\epsilon}{2+\epsilon}\beta_0 .
\end{equation}
Here $x$ and $y$ are defined as in
Section~\ref{sec:isotropic}, while $\kappa$ is given in
equation~(\ref{eq:kappa}) and
\begin{equation}
  \label{eq:aniso:gamma:b}
  \gamma_b = \alpha + \frac{\epsilon}{2}.    
\end{equation}
Equation~(\ref{eq:aniso:b}) is thus the generalisation of
equation~(\ref{eq:iso:b}) for arbitrary $\epsilon$ and $\beta$ linearly
dependent on $\gamma$ (for $\epsilon=3$ and $\beta_0=b=0$, the two equations are
identical). Note that $b<(2+\epsilon)/2\epsilon$ is required to keep the r.h.s
of equation (\ref{eq:aniso:b}) non-negative.

Another differentiation of equation~(\ref{eq:aniso:b}) with respect to $\ln x$
yields the generalisation of equation~(\ref{eq:iso:c})
\begin{eqnarray}
  \gamma^{\prime\prime} &-& \frac{6-\epsilon}{\epsilon}\,\gamma^\prime\,
  \left(\gamma - \frac{2\gamma_a+2\gamma_b-(\epsilon-2)\gamma_1}{6-\epsilon}
  \right)
  \nonumber \\
  \label{eq:aniso:c}
  & = & \frac{2(\epsilon-2)}{\epsilon^2}(\gamma-\gamma_a)(\gamma-\gamma_b)
  (\gamma-\gamma_1)
\end{eqnarray}
with
\begin{equation}
  \label{eq:aniso:gamma:one}
  \gamma_1 = \frac{2(\epsilon-\alpha)}{\epsilon-2}.  
\end{equation}
Again, the singular function $\rho\propto r^{-\gamma_1}$ is a solution, also
identified by \cite{Hansen2004}. As long as
\begin{equation}\label{eq:eps:constraint}
  \epsilon > 2
\end{equation}
to avoid fundamental changes in sign, the only differences between equations
(\ref{eq:aniso:c}) and (\ref{eq:iso:c}) are in numerical values of constants
including $\gamma_a$, $\gamma_b$, and $\gamma_1$.  The topology of the
$(\gamma,\gamma^\prime)$ phase-space for general $\epsilon > 2$ with anisotropy
parametrised as in (\ref{eq:beta:linear}) is thus the same as for the isotropic
case with $\epsilon=3$, and mathematically the solution spaces of the two
problems are isomorphic.  Hence, most of the discussion around
Figure~\ref{fig:flow} carries over here. In particular, for any given
$\epsilon>2$ and $\beta_0\le1$, the fixed points of equation (\ref{eq:aniso:b})
at $\gamma_a$ and $\gamma_b$ satisfy $\gamma_a<\gamma_b$, and they bracket the
third fixed point at $\gamma_1$ as long as
\begin{equation}
  \label{eq:aniso:bounds}
  3-\frac{\epsilon}{2} < \alpha <
  \frac{2+\epsilon}{2} - \frac{\epsilon-2}{2}\,\beta_0\ .
\end{equation}
As before, there then exists a critical value of $\alpha$ for which the three
fixed points are equally spaced: with the requirement $\gamma_1=
\tfrac{1}{2}(\gamma_a+\gamma_b)$, equations (\ref{eq:aniso:gamma:a}),
(\ref{eq:aniso:gamma:b}), and (\ref{eq:aniso:gamma:one}) give
\begin{equation}
  \label{eq:aniso:alpha:crit}
  \alpha_{\mathrm{crit}}=
  \frac{(10-\epsilon)(2+\epsilon)}{2(6+\epsilon)}-
  \frac{2(\epsilon-2)}{6+\epsilon}\beta_0\ ,
\end{equation}
which notably does not depend on the slope $b$ in the linear $\beta$-$\gamma$
relation of equation (\ref{eq:beta:linear}). With $\epsilon=3$ and $\beta_0=0$,
we have $\alpha_{\mathrm{crit}}=35/18$ as in the previous section.

The generic characteristics of the various of solutions to equation
(\ref{eq:aniso:c}) are again determined by whether
$\alpha<\alpha_{\mathrm{crit}}$, $\alpha=\alpha_{\mathrm{crit}}$, or
$\alpha>\alpha_{\mathrm{crit}}$ (see Figure \ref{fig:flow}). For any
\emph{constant} anisotropy ($\beta_0\le 1$ and $b=0$ in
eq.~[\ref{eq:beta:linear}]), the division between physical and unphysical
solutions in all three cases is exactly analogous to the isotropic
specialisation of Section~\ref{sec:isotropic}.  When a gradient in $\beta(r)$ is
allowed, however, the number of physical solutions to the problem becomes
smaller, because $\beta$ is bounded above by 1 at all radii.  In particular, for
the most relevant case of $b>0$ (corresponding, for realistic halo models, to an
outwards increasing radial anisotropy), $\gamma$ must also be bounded above, and
solutions with an outer truncation to the density profile ($\gamma\to\infty$)
are no longer viable.  As a consequence, the only physically possible solutions
with a shallow density cusp in the centre and monotonically increasing slope
$\gamma$ are the analytical solutions that occur only for $\alpha =
\alpha_{\mathrm{crit}}$.

Setting $\alpha=\alpha_{\mathrm{crit}}$ in equation (\ref{eq:aniso:c}) reduces
it to
\begin{equation}\label{eq:aniso:d}
  \gamma^{\prime\prime} -
    \frac{6-\epsilon}{\epsilon}\,\gamma^{\prime}\left(\gamma-\gamma_1\right) = 
      \frac{2(\epsilon-2)}{\epsilon^2}\,
        (\gamma-\gamma_a)(\gamma-\gamma_b)(\gamma-\gamma_1)
\end{equation}
with
\begin{subequations}\label{eq:aniso:gamma:eps}
  \begin{eqnarray}
    \label{eq:aniso:gamma:a:eps}
    \gamma_a(\alpha_{\mathrm{crit}})
      & = & \frac{10-\epsilon}{6+\epsilon}
            + \frac{2(\epsilon+2)}{6+\epsilon}\,\beta_0 ,
    \\
    \label{eq:aniso:gamma:one:eps}
    \gamma_1(\alpha_{\mathrm{crit}})
      & = & \frac{10+3\epsilon}{6+\epsilon}
            + \frac{4}{6+\epsilon}\,\beta_0 ,
    \\
    \label{eq:aniso:gamma:b:eps}
    \gamma_b(\alpha_{\mathrm{crit}})
      & = & \frac{10+7\epsilon}{6+\epsilon}
            - \frac{2(\epsilon-2)}{6+\epsilon}\,\beta_0 .
  \end{eqnarray}
\end{subequations}
Generalising equation (\ref{eq:iso:K}), any solution to equation
(\ref{eq:aniso:d}) conserves a first integral
\begin{equation}\label{eq:aniso:K}
   K = 
   \Big(\gamma^\prime+\tfrac{\epsilon-2}{2\epsilon}
   (\gamma-\gamma_a)(\gamma-\gamma_b)\Big)\,
   \Big(\gamma^\prime-\tfrac{2}{\epsilon}
   (\gamma-\gamma_a)(\gamma-\gamma_b)\Big)^{4/(\epsilon-2)} .
\end{equation}
Thus, for $K=0$ in particular two simple solutions exist.  One is
$\gamma^{\prime} = \gamma_{\mathrm{min}}^{\prime}=
\tfrac{2}{\epsilon}(\gamma-\gamma_a)(\gamma-\gamma_b)$, which defines the
parabola in $(\gamma,\gamma^{\prime})$ phase-space within which all physical
solutions must lie (cf.~Figure \ref{fig:flow}). The other is
\begin{equation} \label{eq:aniso:solution}
  \gamma^\prime = 
      -\frac{\epsilon-2}{2\epsilon}\,(\gamma-\gamma_a)(\gamma-\gamma_b) ,
\end{equation}
which is easily integrated to give an analytic expression for $y(x)\propto
\rho(r)$.\footnote{Note that in the limit $\epsilon=2$, equation
  (\ref{eq:aniso:solution}) is simply $\gamma^{\prime}=0$, while equation
  (\ref{eq:aniso:alpha:crit}) gives $\alpha_{\mathrm{crit}}=2$ for any $\beta_0$
  so the the right-hand side of equation (\ref{eq:aniso:b}) becomes a
  non-negative constant. Therefore any of a \emph{continuum} of singular
  solutions, $y=x^{-\gamma_s}$ for $\gamma_s$ a constant in the interval
  $\gamma_a\le \gamma_s \le \gamma_b$, is a valid critical-$\alpha$ solution for
  $\epsilon=2$. This is the situation that \cite{Hansen2004} is strictly
  relevant to, although pure power-law density profiles such as these are not
  applicable to simulations of dark-matter haloes.} Before writing down this and
other quantities, it proves useful to introduce the auxiliary parameter
\begin{subequations}
  \label{eq:aniso:defs}
  \begin{equation}
    \label{eq:aniso:eta}
    \eta \equiv
        \frac{\epsilon-2}{2\epsilon}(\gamma_b-\gamma_a)  = 
        2\frac{(\epsilon-2)(2-\beta_0)}{6+\epsilon} ,
  \end{equation}    
  the meaning of which becomes clear below. Then
  \begin{equation}
    \label{eq:aniso:epsilon}
    \epsilon = 2\frac{4+3\eta-2\beta_0}{4-\eta-2\beta_0}\ ,
  \end{equation}
  so we have from equation (\ref{eq:aniso:alpha:crit}) that
  \begin{equation}
    \label{eq:aniso:alpha:crit:eta}
    \alpha_{\mathrm{crit}} =
    \eta+2 - \frac{4\eta}{4-\eta-2\beta_0}\ ,
  \end{equation}
  and from equations (\ref{eq:aniso:gamma:eps}),
  \begin{eqnarray}
    \label{eq:aniso:gamma0}
    \gamma_0 \equiv \gamma_a(\alpha_{\mathrm{crit}}) 
      & = & 1-\eta/2+\beta_0,
      \\
    \label{eq:aniso:gamma1}
    \gamma_1(\alpha_{\mathrm{crit}})
      & = & 2+\eta/4+\beta_0/2,
      \\
    \label{eq:aniso:gammai}
    \gamma_\infty \equiv \gamma_b(\alpha_{\mathrm{crit}})
      & = &   3 + \eta .
  \end{eqnarray}
\end{subequations}
Integrating equation~(\ref{eq:aniso:solution}), we then find for the density
\begin{equation}
  \label{eq:aniso:rho}
  \rho(r) \propto x^{-\gamma_0}(1+x^\eta)^{-(\gamma_\infty-\gamma_0)/\eta}.
\end{equation}
Thus, the parameter $\eta$ governs the speed of the transition between the
power-law asymptotes $\rho\propto r^{-\gamma_0}$ at small radii and $\rho\propto
r^{-\gamma_\infty}$ at large radii.\footnote{These solutions are members of the
  much broader class of `$\alpha\beta\gamma$' models discussed by
  \cite{Zhao1996}. In \citeauthor{Zhao1996}'s notation (which is completely
  different from ours), the profiles of equation (\ref{eq:aniso:rho}) have
  $(\alpha, \beta, \gamma) = (\eta^{-1}, \gamma_\infty, \gamma_0) = (\eta^{-1},
  3+\eta, 1-\eta/2+\beta_0)$.}  In addition, the constant $\beta_0$ in equation
(\ref{eq:beta:linear}) takes on physical meaning as the velocity anisotropy at
the centre of the density distribution.

Astonishingly, the gradient $b\equiv\mathrm{d}\beta/\mathrm{d}\gamma$ in
equation~(\ref{eq:beta:linear}) does not appear in equation~(\ref{eq:aniso:c})
or any subsequent relations, implying that the influence of anisotropy on the
density profile is entirely determined by the situation in the centre, \emph{if}
$\beta$ depends linearly on $\gamma$ as we have assumed. For any fixed $\epsilon
> 2$, the main effect of a radially biased velocity ellipsoid at the centre
($0<\beta_0\le1$) is to steepen the inner power law $\gamma_0$ relative to its
isotropic value, and make $\gamma_\infty$ smaller, i.e., the outer density
profile shallower. The reverse holds for a tangential anisotropy, $\beta_0<0$.
In order to keep $\gamma_0\ge0$ then (so the density does not decrease towards
$r\to0$), we require
\begin{equation}
  \beta_0 \ge (\epsilon-10)/(4+2\epsilon)\ ,
\end{equation}
which excludes very strong tangential biases. Isotropic models are allowed only
for $\epsilon\le10$, a limit which is not relevant to dark-matter haloes, but
corresponds to the classic \cite{Plummer1911} sphere. On the other hand, the
physical requirement $\beta_0\le1$ implies $\gamma_0 < 2$ and $\gamma_\infty >
3$ for any $\epsilon > 2$. As a result, all the critical-$\alpha$ density
profiles in equation~(\ref{eq:aniso:rho}) have a finite total mass, and fully
analytical $\sigma_r^2(r)$ and $M(r)$ profiles which are of the same basic form
as in equations~(\ref{eq:iso:model}).

To obtain these profiles in detail, we first express the linear relation between
$\beta$ and $\gamma$ in terms of $\beta_\infty = \lim_{r\to\infty}\beta(r)$
instead of $b$, namely,
\begin{equation}
  \label{eq:beta:gamma}
  \beta(r) = \beta_0 + \frac{\beta_\infty-\beta_0}{\gamma_\infty-\gamma_0}\,
  \big(\gamma(r)-\gamma_0\big).
\end{equation}
Then it is straightforward to show that for $\alpha=\alpha_{\mathrm{crit}}$, 
\begin{subequations}
  \label{eq:aniso:model}
  \begin{eqnarray}
    \label{eq:aniso:kappa}
    \kappa &=& 
    \frac{1}{8}(4+\eta-2\beta_0)(4+\eta-2\beta_\infty),
    \\[0.5ex]
    \label{eq:aniso:density}
    \rho(r) &=& \frac{4+\eta-2\beta_0}{8\pi}\frac{M_{\mathrm{tot}}}{r_0^3}
    x^{-\gamma_0} \big(1+x^\eta\big)^{-(\gamma_\infty-\gamma_0)/\eta}.
    \\[0.5ex]
    \label{eq:aniso:gamma}
    \gamma(r) &=& \frac{\gamma_0+\gamma_\infty x^\eta}{1 + x^\eta},
    \\[0.9ex]
    \label{eq:aniso:beta}
    \beta(r) &=& \frac{\beta_0 + \beta_\infty x^\eta}{1+x^\eta},
    \\[0.5ex]
    \label{eq:aniso:sigmar}
    \sigma_r^2(r) &=&
    \frac{1}{4+\eta-2\beta_\infty}\frac{G M_{\mathrm{tot}}}{r_0}
    x^{-1}\left(\frac{x^\eta}{1+x^\eta}\right)^
    {\tfrac{\gamma_\infty-\gamma_0}{\eta}-2},
    \\[0.5ex]
    \label{eq:aniso:mass}
    M(r) &=& M_{\mathrm{tot}}
    \left(\frac{x^\eta}{1 + x^\eta}\right)^
    {\tfrac{\gamma_\infty-\gamma_0}{\eta}-1},
    \\[0.5ex]
    \label{eq:aniso:vcirc}
    V_c^2(r) &=& \frac{G M_{\mathrm{tot}}}{r_0} x^{-1}
    \left(\frac{x^\eta}{1+x^\eta}\right)^
    {\tfrac{\gamma_\infty-\gamma_0}{\eta}-1},
  \end{eqnarray}
  where $M_{\mathrm{tot}}$ is related to $\rho_0$ by the condition
  $\rho(r=r_0)=\rho_0$ and $\eta$, $\gamma_0$, and $\gamma_\infty$ are, of
  course, given in terms of $\epsilon$ and $\beta_0$ by
  equations~(\ref{eq:aniso:defs}) and satisfy $(\gamma_\infty-\gamma_0)/\eta=
  2\epsilon/(\epsilon-2)$. The total one-dimensional velocity-dispersion
  profile in these models is also analytical, being given simply by
  $\sigma_{\mathrm{tot}}^2(r)=\sigma_r^2(r)(1-2\beta(r)/3)$.  Note that for
  $\epsilon=3$ and $\beta_0=\beta_\infty=0$ (the specialised case considered in
  Section~\ref{sec:isotropic}), $\eta=4/9$ and we recover all of equations
  (\ref{eq:iso:model}). As in that case, the velocity-dispersion and
  circular-velocity profiles show peaks in this more general situation:
  $\sigma_r^2$ has its maximum at $x=(1-\eta/2-\beta_0)^{1/\eta}$ and $V_c^2$,
  at $x=(1+\eta/2-\beta_0)^{1/\eta}$.  Finally, the gravitational potential is
  \begin{eqnarray}
    \label{eq:aniso:pot}
    \Phi(r) &=& -\frac{G M_{\mathrm{tot}}}{\eta r_0}
    B_{\frac{1}{1+x^\eta}}\left(\tfrac{1}{\eta},
      \tfrac{1-\beta_0}{\eta}+\tfrac{1}{2}\right)
    \\[0.5ex]
    \label{eq:aniso:pot:alt}
    &=& -\frac{G M_{\mathrm{tot}}}{\eta r_0}
    \left[B\left(\tfrac{1}{\eta},\tfrac{1-\beta_0}{\eta}+\tfrac{1}{2}\right)
      -
    B_{\frac{x^\eta}{1+x^\eta}}\left(
      \tfrac{1-\beta_0}{\eta}+\tfrac{1}{2},\tfrac{1}{\eta}\right)
    \right]
  \end{eqnarray}
  where again $B_u(p,q)\equiv\int_0^u t^{p-1}(1-t)^{q-1}\mathrm{d}t$ is the
  incomplete beta function and $B(p,q)\equiv
  B_1(p,q)=\Gamma(p)\Gamma(q)/\Gamma(p+q)$ the (complete) beta function. In the
  limit of large radii $\Phi\to-GM_{\mathrm{tot}} r^{-1}$, while for small radii
  \begin{equation}
    \label{eq:aniso:pot:asym}
    \Phi(r) \approx -\frac{G M_{\mathrm{tot}}}{r_0}
    \left[\tfrac{1}{\eta}
      B\left(\tfrac{1}{\eta},\tfrac{1-\beta_0}{\eta}+\tfrac{1}{2}\right)
      - \tfrac{2}{2+\eta-\beta_0} x^{\eta/2+1-\beta_0} \right].
  \end{equation}
\end{subequations}
It is worth noting that at small radii $\rho\propto r^{-\gamma_0}$ and
$\sigma_r^2\propto r^{\gamma_0-2\beta_0}$. Thus the pressure $\rho\sigma_r^2\to
r^{-2\beta_0}$, which diverges for radial anisotropies at the centre
($\beta_0>0$) but approaches a constant for central isotropy and vanishes for
tangentially biased central velocity distributions.

Aside from its pleasing---and somewhat surprising---simplicity in the face of a
nontrivial radial variation of $\beta(r)$, the family of models defined by
equations~(\ref{eq:aniso:model}) is further interesting because the linear
`$\beta$-$\gamma$ relation' in equation~(\ref{eq:beta:linear}) or
(\ref{eq:beta:gamma}) is precisely what \cite{HansenMoore2005} have suggested is
a generic result of collisionless collapses, mergers, and relaxation processes
(see their Figure 2).

Moreover, it is beneficial that for any $\epsilon$ these anisotropic models have
the same critical value of the exponent $\alpha$ in our $\rho$-$\sigma$ relation
(\ref{eq:alpha}) as do the fully isotropic models, just so long as isotropy
holds at the centre alone ($\beta_0=0$). Simulated dark-matter haloes indeed
tend to be roughly isotropic at their centres and radially anisotropic in their
outer parts. Thus, as was discussed at the end of Section~\ref{sec:isotropic},
it is again striking that the data shown in the right-hand panels of Figure
\ref{fig:rhosig} exhibit a scaling $\rho/\sigma_r^3\sim r^{-\alpha}$ with
$\alpha\simeq1.92$ very close to the expected critical value
($35/18=1.9\overline{4}$) for $\epsilon=3$ and $\beta_0=0$.

With these points in mind, we now go on to a detailed fitting of our models in
equations~(\ref{eq:aniso:model}) to the density and velocity-dispersion profiles
of dark-matter haloes simulated by \cite{DiemandMooreStadel2004a,
  DiemandMooreStadel2004b}.

\section{Comparison with simulated haloes}
\label{sec:simhalo}

In order to compare the anisotropic models in equation (\ref{eq:aniso:model})
against numerical dark-matter haloes, we again make use of the simulations
published by \cite{DiemandMooreStadel2004a,DiemandMooreStadel2004b}, which we
referred to in Section~\ref{sec:intro} (Fig.~\ref{fig:rhosig}). To repeat, these
include four galaxy-sized haloes (virial masses $1 - 2 \times10^{12}\,M_\odot$)
and six clusters (virial masses $2.4\times10^{14}\,M_\odot -
1.3\times10^{15}\,M_\odot$). All were evolved to redshift $z=0$ except for two
clusters which were run further ahead in time to complete mergers; see
\citeauthor{DiemandMooreStadel2004a} for full details.

In addition to density profiles, the data on these haloes include the enclosed
mass profiles $M(r)$ and the radial and tangential components of velocity
dispersion, $\sigma_r^2(r)$ and $\sigma_t^2(r) =
\sigma_\theta^2(r)+\sigma_\phi^2(r)$. From the run of $\rho(r)$ in a series of
spherical shells $\{r_i\}$, we have estimated the local density gradient as
$\gamma(r_i)=-[\log\,\rho(r_{i+1})-\log\rho(r_{i-1})] /
[\log\,r_{i+1}-\log\,r_{i-1}]$. From the velocity dispersions, we have
calculated $\beta(r_i)=1-\sigma_t^2(r_i)/2\sigma_r^2(r_i)$ and the total
one-dimensional dispersion, $\sigma_{\mathrm{tot}}^2(r_i) = \tfrac{1}{3}
[\sigma_r^2(r_i)+\sigma_t^2(r_i)]$.

Five parameters define the models in equations~(\ref{eq:aniso:model}): the
normalisation constant $M_{\mathrm{tot}}$ and the scale radius $r_0$; the
parameters $\epsilon$ and $\beta_0$, which fix $\eta$, $\gamma_0$, and
$\gamma_\infty$ (and thus the shapes of all profiles); and finally
$\beta_\infty$, the velocity anisotropy at $r\to\infty$. Together these must
suffice to describe the separate $\rho(r)$, $\sigma_r^2(r)$, and $\beta(r)$
profiles for a dark-matter halo. While it is of course possible to fit each of
the ten simulated haloes individually, we are more interested here in the
question of whether some `universal' parameter values might apply.

\begin{figure*}
  \centerline{\hfil
    \resizebox{135mm}{!}{\includegraphics{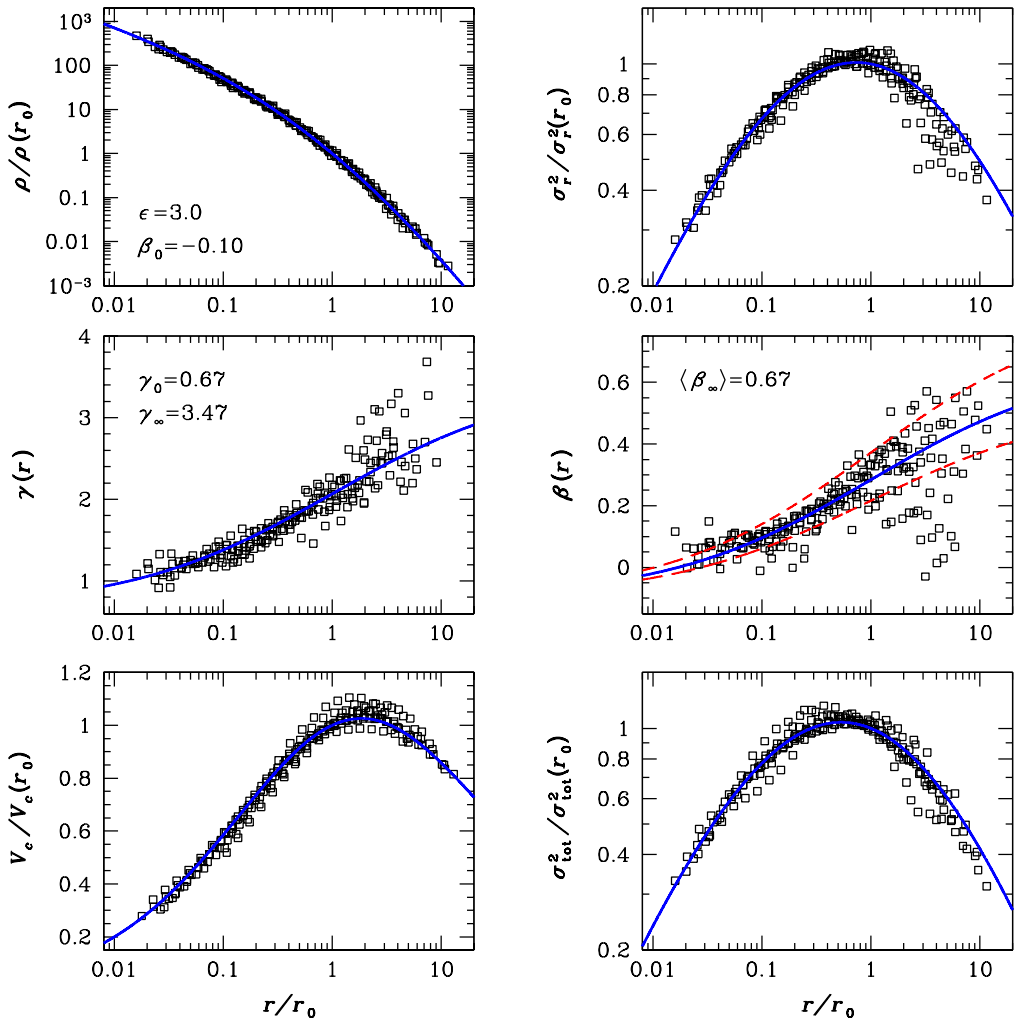}}\hfil
  }
  \caption{
    \label{fig:simhalo}
    Comparison of our analytical, 'critical-$\alpha$' models in equation
    (\ref{eq:aniso:model}) with the same ten simulated CDM haloes from
    \citet{DiemandMooreStadel2004a,DiemandMooreStadel2004b} that have already
    been employed in Fig.~\ref{fig:rhosig}. The exponent $\epsilon$ in the
    $\rho$-$\sigma$ relation (\ref{eq:alpha}) was kept fixed at $\epsilon=3$,
    while a \emph{global} $\beta_0$ was fit to the density and velocity
    dispersion profiles of all haloes simultaneously. For each halo, \emph{
      individual} values for the scale radius $r_0$, normalisation constant
    $M_{\mathrm{tot}}$, and asymptotic anisotropy $\beta_\infty$ at $r\to\infty$
    have been fitted (see text for more details).  The \emph{solid} curves show
    the dimensionless model profiles given the global $\epsilon$ and $\beta_0$.
    Squares represent the data from the simulated haloes, with the various
    profiles for each halo scaled by their best-fit values at the fitted $r_0$
    for that particular halo.  For the anisotropy profile, the \emph{solid}
    curve represents a mean model, in the sense of a mean value for
    $\beta_\infty$, while the \emph{dashed} curves indicate the range of
    best-fit values for $\beta_\infty$.  }
\end{figure*}
Thus, we start by assuming that all haloes follow the same $\rho$-$\sigma$
relation, $\rho/\sigma_r^\epsilon \propto r^{-\alpha_{\mathrm{crit}}}$, with a
single value of $\epsilon$. That this is likely so is already suggested by
Figure~\ref{fig:rhosig}. We further assume that the central velocity anisotropy
$\beta_0$ is the same for all haloes. This is more of a debatable
contention---even though most simulations show similar (low) levels of
anisotropy in their innermost resolved regions, there is no clear evidence that
$\beta_0$ always tends to a single value. Nevertheless, if $\epsilon$ and
$\beta_0$ both are `universal' then the shape of the density profile must be too
in these models, and so this possibility is worth examining.

We proceed by defining a grid of ($\epsilon, \beta_0$) values. For each pair in
turn, we compute $\eta$, $\gamma_0$, and $\gamma_\infty$ from
equations~(\ref{eq:aniso:defs}), which allows for the calculation of
dimensionless model profiles $\rho_{\mathrm{mod}}(r)$ and
$\sigma_{r,\mathrm{mod}}^2(r)$. For each of the ten haloes, we then find values
for $M_{\mathrm{tot}}^{(j)}$, $r_0^{(j)}$, and $\beta_\infty^{(j)}$ to minimise
the sum of absolute deviations,
\begin{equation}\label{eq:dev}
  \Delta_j(\epsilon,\beta_0) = \sum_{i=1}^{N_j} \left\{ 
   \left|\log\frac{\rho_{\mathrm{mod}}(r_i)}{\rho(r_i)}\right|
   + \epsilon
   \left|\log\frac{\sigma_{r,\mathrm{mod}}(r_i)}{\sigma_r(r_i)}\right|
   \right\},
\end{equation}
which is more robust against outliers than the standard $\chi^2$ statistic.
Here $N_j$ is the number of data points in the $j$th halo; typically, $N_j=21$.
The total deviation that is minimised for each ($\epsilon, \beta_0$) is thus
\begin{equation}\label{eq:devtot}
  \Delta_{\mathrm{tot}}(\epsilon, \beta_0) = \sum_{j=1}^{N_{\mathrm{halo}}}
    \Delta_j (\epsilon,\beta_0)\ .
\end{equation}
Ultimately, we find the set $\{\epsilon, \beta_0; M_{\mathrm{tot}}^{(j)},
r_0^{(j)}, \beta_\infty^{(j)}, j=1, \dots, 10\}$ for which
$\Delta_{\mathrm{tot}}$ is the minimum over our original grid. Strictly
speaking, this occurs at $\epsilon=3.2$ and $\beta_0=-0.05$, but the minimum is
rather shallow and the best fit with $\epsilon=3$ exactly (which requires
$\beta_0=-0.10$) is not significantly worse. The details of this latter fit are
shown in Figure \ref{fig:simhalo}.

The upper panels of Figure \ref{fig:simhalo} show the density and (radial)
velocity-dispersion data that were used to constrain the model parameters. To
emphasise the shapes of the distributions, the radial coordinate in each halo
$j$ has been normalised by the fitted value of $r_0^{(j)}$, and the densities
and velocity dispersions have been normalised by their fitted values at
$r_0^{(j)}$. The best-fitting dimensionless profiles, from equations
(\ref{eq:aniso:density}) and (\ref{eq:aniso:sigmar}), are drawn as the
\emph{bold} curves. The rms relative deviations from these curves, $(\rho -
\rho_{\mathrm{mod}})/\rho$ and $(\sigma_r^2 - \sigma_{r,
  \mathrm{mod}}^2)/\sigma_r^2$, are both of order 11\%--12\% for all ten haloes
combined.

The left-middle panel of Figure \ref{fig:simhalo} shows the negative logarithmic
density gradient $\gamma(r)$ as a function of $r/r_0^{(j)}$ in the simulated
haloes against the model curve given by equation (\ref{eq:aniso:gamma}) for
$\epsilon=3$ and $\beta_0=-0.10$. Given these parameters,
equations~(\ref{eq:aniso:defs}) imply that the power-law slopes at $r\to0$ and
$r\to\infty$ are $\gamma_0=2/3$ and $\gamma_\infty=52/15=3.4\overline{6}$,
respectively. The transition from the inner to the outer power law is rather
gradual, with $\eta=7/15=0.6\overline{6}$. Recall that $r_0$ is defined as the
radius at which $\gamma(r_0)=\gamma_1$, with
$\gamma_1=\tfrac{1}{2}(\gamma_0+\gamma_\infty)=31/15$ in this case.

The right-middle panel then shows the `observed' anisotropy parameter $\beta(r)$
vs.~the scaled radius $x=r/r_0^{(j)}$. The \emph{bold} curve traces the model
relation~(\ref{eq:aniso:beta}) for $\beta_0=-0.10$ and $\beta_\infty=0.67$,
which is the average of the ten different $\beta_\infty^{(j)}$ values obtained
by fitting the $\sigma_r^2$ profile of each halo separately. The \emph{dashed}
curves have the same $\beta_0=-0.10$ but $\beta_\infty=0.53$ and
$\beta_\infty=0.84$, corresponding to the minimum and maximum of the fits to the
ten haloes.  Evidently, these curves together account for much of the observed
scatter in $\beta(r)$. Note that the average $\beta_\infty$ value implies a
slope for the $\beta$-$\gamma$ relation~(\ref{eq:beta:gamma}) of
$(\beta_\infty-\beta_0)/(\gamma_\infty-\gamma_0)\simeq0.28$. This is comparable
to the slope $\mathrm{d}\beta/\mathrm{d}\gamma\approx 0.19$ inferred by
\cite{HansenMoore2005} from their investigations of a completely different set
of simulated haloes.  However, it should also be noted that
\citeauthor{HansenMoore2005} found a relatively tight correlation between
$\beta$ and $\gamma$ (see their Figure 2).  Comparison of the data points in the
two middle panels of our Figure \ref{fig:simhalo} shows somewhat more scatter in
any empirical $\beta$-$\gamma$ relation for the haloes we are working with.
This is particularly evident at relatively large radii, $r/r_0\ga 2$. In fact,
it is not clear that a `universal' slope in a linear $\beta$-$\gamma$
relationship (i.e., a unique value of $\beta_\infty$) can describe all of these
data if the density profile is strictly 'universal' (i.e. if there truly is a
single value for $\beta_0$ as well as $\epsilon$).

The bottom two panels of Figure \ref{fig:simhalo} complete the comparison with
our models. The left panel shows the re-normalised circular-velocity profile,
$\sqrt{M(r)r_0/M(r_0)r}$ vs.~$r/r_0$, which is essentially equivalent to the
top-left panel showing the model and observed density profiles. The right-hand
panel shows the normalised total one-dimensional velocity dispersion profile,
which is obtained from the two panels above it: $\sigma_{\mathrm{tot}}^2(r)=
\sigma_r^2(r)(1-2\beta(r)/3)$. Note that the scatter of
$\sigma_{\mathrm{tot}}^2$ about our model is some 25\% smaller than that of
$\sigma_r^2$: the scatter away from the model curve in the upper-right panel is
compensated by the scatter of $\beta(r)$ in the middle-right panel.

\begin{figure} \label{fig:resids}
  \centerline{\hfil
    \resizebox{81mm}{!}{\includegraphics{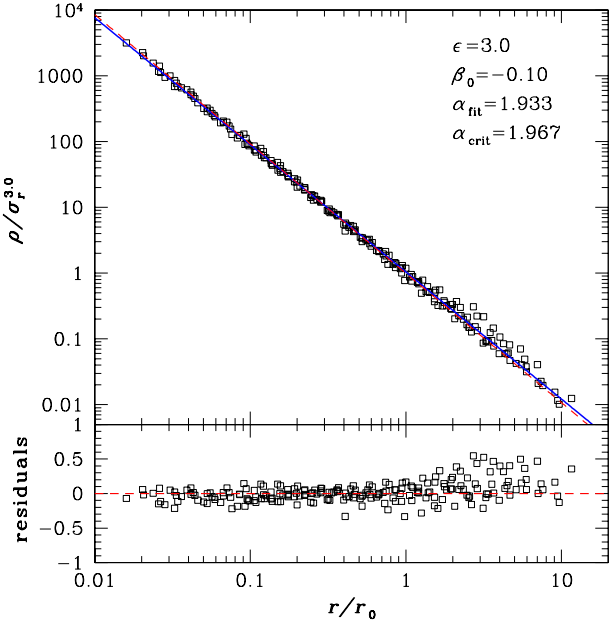}}
  }
  \caption{
    \emph{Top panel}: plot of $\rho/\sigma_r^3$ against radius for the ten
    simulated CDM haloes already employed in Figs.~\ref{fig:rhosig} and
    \ref{fig:simhalo}.  In fact, this plot is very similar to the right panel of
    Fig.~\ref{fig:rhosig}, the only distinction being that we use our best-fit
    values for individual scale radii $r_0^{(j)}$ to normalise the data in each
    halo.  The \emph{solid} line gives the best-fit power-law relation, with
    exponent $\alpha_{\mathrm{fit}}=1.933$. The \emph{dashed} line is the power
    law implied by the fit of our critical-$\alpha$ model to the haloes' density
    and velocity-dispersion profiles in Figure \ref{fig:simhalo}: with
    $\epsilon=3$ and $\beta_0=-0.10$, $\alpha_{\mathrm{crit}}=1.9\overline{6}$.
    \emph{Bottom panel}: Relative residuals of $(\rho/\sigma_r^3)$ about the
    latter, critical-$\alpha$ power law assumed by our model fit.  }
\end{figure}

Overall, it is surprising how well our simple model is able to reproduce the
main features of the spatial structure \emph{and the kinematics} of these
simulated dark-matter haloes. However, it remains to be checked that the
$\rho$-$\sigma$ relation in the `observed' haloes is consistent with that
required by the analytical models we have fit. Specifically, by using
equations~(\ref{eq:aniso:model}) we have assumed that the value of $\alpha$ is
the critical one given by equation (\ref{eq:aniso:alpha:crit}). For $\epsilon=3$
and $\beta_0=-0.10$ as in Figure \ref{fig:simhalo}, this is
$\alpha_{\mathrm{crit}}= 59/30=1.9\overline{6}$.  Figure \ref{fig:resids} plots
the $\rho/\sigma_r^3$ data points from the ten
\citeauthor{DiemandMooreStadel2004a}~haloes (normalised by our fitted values of
$\rho_0^{(j)}$ and $\sigma_{r,0}^{(j)}$ for each halo individually) against the
scaled radius $r/r_0^{(j)}$. The \emph{dashed} line has slope
$-\alpha_{\mathrm{crit}}$ and gives a reasonable description of the data. In
fact, direct linear regression (with a 3-$\sigma$ clipping applied) yields a
best-fit slope of $-1.933$, only$\simeq$1.5\% different from the expected
value.\footnote{The fitted slope here differs slightly from that in the
  right-hand panel of Fig.~\ref{fig:rhosig} because now we have scaled to the
  fitted radius $r_0^{(j)}$ in each halo, rather than to model-independent, but
  cruder, estimates of $r_2$.}  This is drawn as the \emph{solid} line in
Fig.~\ref{fig:resids}. The relative residuals from the critical-$\alpha$ model
line, $[(\rho/\sigma_r^3)-(\rho/\sigma_r^3)_{\mathrm{mod}}]/(\rho/\sigma_r^3)$,
are shown in the lower panel. Evidently, the largest deviations from this power
law occur at $r/r_0\ga 2$, which is where the $\sigma_r^2(r)$ and $\beta(r)$
profiles scatter most in Figure \ref{fig:simhalo}.

As a last test, we wish to compare our model against the estimated density
profile \emph{only} of an extremely high-resolution halo simulated by
\cite{DiemandEtal2005}. This is again a cluster-sized halo. As
\citeauthor{DiemandEtal2005} describe in detail, this particular halo was
defined by first performing a simulation with spatial resolution of $10^{-3}$ in
units of the virial radius, evolving the run to a high redshift.  The most
central part of the density profile was then scaled to match onto the outer
parts of a lower-resolution halo previously evolved to $z=0$. Because of this
estimation procedure, we do not have the full velocity-dispersion and anisotropy
profiles for this cluster.

Even in the absence of kinematical data, we can fit for all of $\epsilon$,
$\beta_0$, $M_{\mathrm{tot}}$, and $r_0$ using the density profile alone. In
practice, we set $\epsilon=3$ and only fit for $\beta_0$ and the two
normalisation factors by minimising the sum of absolute deviations
$\Delta=\sum_{i} \left|\log\,\rho_{\mathrm{mod}}(r_i)-\log\,\rho(r_i)\right|$.
In this case, the shape of the density profile at small $r$ requires a slightly
larger $\beta_0$ than we found for Figure \ref{fig:simhalo}: $\beta_0\simeq0.03$
(which, reasonably, is still nearly isotropic). This is one indication that
dark-matter halo density profiles may not be exactly universal after all, even
if the value of $\epsilon$ is. Some of the scatter found in $\gamma$ at the
resolution limit of numerous simulations \citep[e.g.,][]{NavarroEtal2004,
  DiemandMooreStadel2004b, FukushigeKawaiMakino2004} may in fact be real and, in
our model at least, connected to non-universal halo \emph{kinematics}.

\begin{figure}
  \centerline{\hfil
    \resizebox{81mm}{!}{\includegraphics{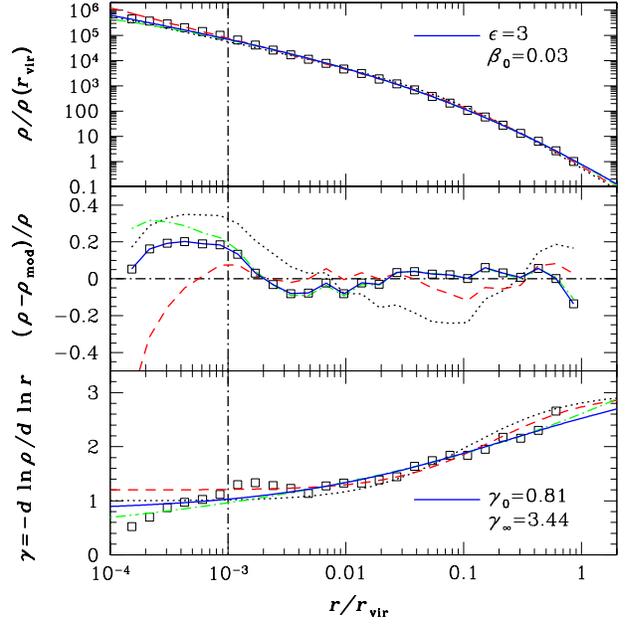}}
  }
  \caption{
    \label{fig:hires}
    Comparison of the best fit (\emph{blue solid curves}) of
    equation~(\ref{eq:aniso:density}) to the high resolution (`billion
    particle') halo of \citet{DiemandEtal2005}. For this fit, $\epsilon=3$ was
    fixed \emph{a priori} and only $\beta_0$, $r_0$, and $M_{\mathrm{tot}}$ were
    allowed to vary. The vertical \emph{dot-dashed} line indicates the
    resolution limit of the simulation, according to
    \citeauthor{DiemandEtal2005}. The \emph{black dotted}, \emph{green
      dash-dot}, and \emph{red long-dashed} curves correspond to the best fits
    of the profiles proposed by, respectively, \citet{NavarroFrenkWhite1996}
    (see eq.~[\ref{eq:nfw96}]), \citet{NavarroEtal2004} (eq.~[\ref{eq:nav04}]),
    and \citet{DiemandEtal2005} (eq.~[\ref{eq:die05}]).  }
\end{figure}

The top panel of Figure \ref{fig:hires} plots the best fit of equation
(\ref{eq:aniso:density}) against the data for this one halo, now with the virial
radius $r_{\mathrm{vir}}$ taken as the normalisation point; this model is given
by the solid (\emph{blue}) curve. For comparison with other fitting functions
employed in the literature, we have also found the best-fitting NFW profile,
\begin{equation}\label{eq:nfw96}
  \rho(r)/\rho_2=4\,(r/r_2)^{-1}(1+r/r_2)^{-2}\ ,
\end{equation}
which we show as the dotted (\emph{black}) curve. The best-fitting function of
the type suggested by \cite{NavarroEtal2004},
\begin{equation}\label{eq:nav04}
  \rho(r)/\rho_2=\exp\left\{-\left(2/\alpha_N\right)
                     \left[\left(r/r_2\right)^{\alpha_N}-1\right]\right\} ,
\end{equation}
is shown as the dash-dot (\emph{green}) curve. Here $\alpha_N$ is a free
parameter in the fit, which we find to be $\alpha_N=0.144$. Finally, the fitting
formula preferred by \cite{DiemandEtal2005},
\begin{equation}\label{eq:die05}
  \rho(r)/\rho_s=2^{1.8}\,(r/r_s)^{-1.2}(1+r/r_s)^{-1.8}\ ,
\end{equation}
is drawn as the long-dashed (\emph{red}) curve. Within the resolved radial range
of the simulation, $r/r_{\mathrm{vir}}\ge10^{-3}$, there is little obvious
difference between these fits in a plot of $\rho$ vs. $r$.

In the middle panel of Figure \ref{fig:hires} we show the fractional residuals
from each of the four models fitted to $\rho(r)$. The points joined by the
\emph{solid blue} curve denote the residuals from the fit of our model;
residuals from the others just listed are in the same line types and colours as
in the upper panel. The NFW model is clearly worse than any of the others, but
the remaining three are competitive. The rms fractional density residual about
the best NFW fit is 16\%; that about the \citeauthor{NavarroEtal2004} function
is 6.3\%; that about the \cite{DiemandEtal2005} formula is 5.2\%; and that about
our model with $\epsilon=3$ and $\beta_0=0.03$ is 6.0\%.  The main source of the
slightly higher global scatter in our model vs.\ the fitting function of
\citeauthor{DiemandEtal2005} is the single density point closest to the
resolution limit of this simulation.

It is noteworthy that our best-fit dynamical model is almost identical to the
best-fitting \citeauthor{NavarroEtal2004} function within the resolved radial
range, as the latter has been shown to provide a very accurate description of
many simulated dark-matter haloes \citep[e.g.,][]{NavarroEtal2004,
  DiemandMooreStadel2004b}.  At some level, it is not surprising that either of
these curves is an improvement over, say, the NFW profile, since the former both
involve three free parameters in $\rho(r)$ (recall that we fixed $\epsilon=3$ in
fitting our model to this halo) while the NFW function contains only two. The
advantage to our model, of course, is that the extra degree of freedom in
fitting the density profile is also used simultaneously to predict the
\emph{anisotropic} kinematics of the halo.  For example, from equation
(\ref{eq:aniso:alpha:crit}) we would predict $\alpha =
\alpha_{\mathrm{crit}}=1.938$ in the $\rho$-$\sigma$ relation (\ref{eq:alpha})
for this halo. Perhaps coincidentally, this is almost exactly the
$\alpha_{\mathrm{fit}}$ found in Figure \ref{fig:resids} for the ten
lower-resolution haloes from \cite{DiemandMooreStadel2004a,
  DiemandMooreStadel2004b}.

Finally, the bottom panel of Figure \ref{fig:hires} shows the negative
logarithmic density slope estimated as a function of radius directly from the
data of \cite{DiemandEtal2005}, against the behaviours predicted by each of the
four fits to the density profile. With $\epsilon=3$ and $\beta_0=0.03$, our
model has $\gamma\to\gamma_0=0.81$ as $r\to0$, and $\gamma\to\gamma_\infty=3.44$
as $r\to\infty$. The rollover to $\gamma_0$ is very gradual however (from
eq.~[\ref{eq:aniso:eta}], $\eta=0.438$ here) and at the resolution limit of the
simulation we still find $\gamma\simeq1$ for the fit. Apparently, still
higher-resolution simulations are required to distinguish clearly between our
density model and others, such as that of \citeauthor{DiemandEtal2005}, with
different asymptotic slopes in the limit $r\to 0$.

\section{Summary and discussion}
\label{sec:summary}

As a rule, investigations into the structure of simulated dark-matter haloes
have focused on the nearly 'universal' shape of their density profiles. This
approach is plagued, however, by the fact that the haloes are only ever resolved
over 2--3 decades in radius. Since the logarithmic density slope clearly varies
smoothly over this entire range, in a way that is not anticipated from any first
principles, it is then difficult to characterise unambiguously the true form of
$\rho(r)$ with necessarily ad hoc, empirical fitting functions. In particular,
the value of the asymptotic central power-law slope (if a single, universal
asymptote even exists in the limit $r\to 0$) remains ill-constrained. Thus, the
starting point of our analysis in this paper was the remarkable, and much
simpler, empirical fact (\citeauthor{TaylorNavarro2001}
\citeyear{TaylorNavarro2001}, see also \citealt{Hansen2004}) that simulated
haloes satisfy, over their whole resolved range, a $\rho$-$\sigma$ relation of
the general form
\[
  \frac{\rho(r)}{\sigma_r^{\epsilon}(r)} \propto r^{-\alpha}
\]
for $\sigma_r$ the radial component of velocity dispersion, and
$\epsilon\simeq3$. (As it happens, using the total one-dimensional velocity
dispersion also leads to a nearly power-law $\rho$-$\sigma$ relation, with a
slightly different value for $\alpha$; see \citeauthor{TaylorNavarro2001}, and
Figure \ref{fig:rhosig} above.) A single power-law dependence such as this is
clearly easier to recognise, and to quantify accurately, than is a
radius-dependent curvature in the density profile. It is also the simplest
nontrivial form that any halo relation could take, and hence the $\rho$-$\sigma$
scaling may be the most fundamental aspect of dark-matter haloes.

We should note here that, while $\rho/\sigma^3$ has the dimension of phase-space
density, its interpretation as phase-space density \citep{TaylorNavarro2001,
  WilliamsEtal2004}, coarse-grained or not, is problematic\footnote{Only if the
  functional form of the distribution of velocities were independent of radius,
  i.e.\ the distribution function separable, would $\rho/\sigma^3$ equal the
  \emph{average} phase-space density at radius $r$. However, there is no a
  priori reason why this situation should be satisfied in dark-matter haloes.}.
On the other hand however, if the phase-space density were in some sense of
power-law form, we might expect $\rho/\sigma^3$ to closely, but not exactly,
follow a power law, too.

Our basic assumption has been that a power-law $\rho$-$\sigma$ relation holds
for all radii, not just those resolved by simulations. It is then possible to
use this as a constraint on the spherical Jeans equation in order to
\emph{derive} the density profile of an equilibrium halo. In
Section~\ref{sec:isotropic} we did just this, for a specialised case with
$\epsilon=3$ fixed and velocity isotropy assumed. This part of the analysis is
thus similar to the original work of \cite{TaylorNavarro2001}, and to aspects of
\citeauthor{WilliamsEtal2004}~(\citeyear{WilliamsEtal2004}, see also
\citealt{Hansen2004}). However, our study has expanded considerably on these
others, as we investigated the full space of all physical solutions to the
isotropic Jeans equation under a power-law $\rho$-$\sigma$ relation with
arbitrary exponent $\alpha$. Most importantly, we have found that only one
solution $\rho(r)$ exists which asymptotes to realistic power-law behaviour in
the limits of both small and large radii. This solution occurs only for a single
critical value of $\alpha$ and, for this one solution only, the density,
velocity-dispersion, and enclosed-mass profiles are all fully analytic.

In Section~\ref{sec:aniso} we extended our analysis by allowing for arbitrary
exponents $\epsilon$ in the $\rho$-$\sigma$ relation and, unlike any other
detailled study before, for velocity anisotropy. We found that the problem
remains analytically tractable for any $\epsilon>2$, only if we adopt a
$\rho$-$\sigma$ relation involving the \emph{radial} component $\sigma_r$ of
velocity dispersion rather than the total (empirically both are equally well
motivated, see Fig.~\ref{fig:rhosig}) \emph{and} if we assume the anisotropy
parameter $\beta(r)\equiv1- \sigma_\theta^2/ \sigma_r^2$ to be either spatially
constant or linearly dependent on the logarithmic density slope
$\gamma(r)\equiv-\mathrm{d}\ln\,\rho/\mathrm{d}\ln\,r$.  Under these reasonable
assumptions, there always exists an exact analogue to the fully analytical
density profile we found in the isotropic, $\epsilon=3$ case. This solution
again occurs for a particular, `critical' $\alpha$ value. Moreover, for the
realistic case of $\beta$ linearly increasing with $\gamma$ (corresponding to
outwards ever more radially biased anisotropy), this is the \emph{only} physical
solution which resembles simulated dark-matter haloes.

This result is gratifying for two reasons. First, it connects neatly with an
entirely independent, empirical finding by \cite{HansenMoore2005}, who have
argued on the basis of a variety of simulations that indeed there exists a
roughly linear relation between $\beta(r)$ and $\gamma(r)$ in collisionless
haloes formed through numerous different processes. Second, it turns out that
the \emph{slope} of this linear `$\beta$-$\gamma$ relation' does not affect the
critical value of $\alpha$, nor does it affect the shape of the equilibrium
$\rho(r)$ profile itself. Rather, both of these things are influenced only by
the \emph{central} anisotropy $\beta_0\equiv\beta(r=0)$. Specifically, for a
$\rho$-$\sigma$ relation with $\epsilon=3$ (which ultimately does appear to be
most appropriate), our solution of the Jeans equation with radially varying
velocity anisotropy requires
\begin{subequations}
  \label{eq:e:3:models}
  \begin{equation}
    \label{eq:e:3:alpha}
    \frac{\rho}{\sigma_r^3} \propto r^{-\frac{35}{18}+\frac{2}{9}\beta_0}
  \end{equation}
  and has density
  \begin{equation}
    \label{eq:e:3:rho}
    \rho(r) \propto x^{-\frac{7+10\beta_0}{9}}
    \left(1+x^{\frac{2(2-\beta_0)}{9}}\right)^{-6},
  \end{equation}
  with $x\equiv r/r_0$, radial velocity dispersion
  \begin{equation}
    \label{eq:e:3:sigma}
    \sigma_r^2(r) \propto 
    x^{\frac{7-8\beta_0}{9}}
    \left(1+x^{\frac{2(2-\beta_0)}{9}}\right)^{-4},
  \end{equation}
  and anisotropy
  \begin{equation}
    \label{eq:e:3:beta}
    \beta(r) \equiv 1-\frac{\sigma_\theta^2(r)}{\sigma_r^2(r)} 
    = \frac{\beta_0+\beta_\infty x^{2(2-\beta_0)/9}}
    {1+x^{2(2-\beta_0)/9}}
  \end{equation}
\end{subequations}
for any $\beta_0\le 1$, $\beta_\infty\le 1$. Thus, the halo density profile has
a shallow power-law cusp, $\rho\to r^{-(7+10\beta_0)/9}$ in the limit $r\to 0$,
and steepens monotonically but slowly to $\rho\to r^{-(31-2\beta_0)/9}$ as
$r\to\infty$. More general results for $\epsilon\ne 3$ in the $\rho$-$\sigma$
relation, including analytical expressions for the velocity-dispersion,
enclosed-mass, and circular-velocity profiles, are given in equations
(\ref{eq:aniso:model}) above (with auxiliary definitions in
eqs.~[\ref{eq:aniso:defs}]).

In Section~\ref{sec:simhalo} we fit our analytical, critical-$\alpha$ density
and velocity-dispersion profiles to those in ten dark-matter haloes simulated by
\cite{DiemandMooreStadel2004a, DiemandMooreStadel2004b}. When modelling these
'data', we assumed that all haloes obey a $\rho$-$\sigma$ relation with a single
value of $\epsilon$ \emph{and} have the same central velocity anisotropy
$\beta_0$, so that the halo density profile is necessarily universal. We were
able to find a good fit to all ten haloes simultaneously for $\epsilon=3$ and
$\beta_0=-0.10$, only slightly different from isotropic; see Figure
\ref{fig:simhalo}. For this combination of parameters, our model requires
$\rho/\sigma_r^3 \propto r^{-1.9\overline{6}}$ for self-consistency, and we
showed in Figure \ref{fig:resids} that this is a satisfactory description of the
data.

In Section~\ref{sec:simhalo} we also fit our model to the density profile of an
extremely high-resolution ('billion-particle') halo simulated by
\cite{DiemandEtal2005}, finding good agreement again with $\epsilon=3$ and a
nearly isotropic $\beta_0=0.03$. In this case the fitted density profile has a
central power-law cusp $\rho\to r^{-0.81}$ as $r\to 0$, which steepens gradually
to $\rho\to r^{-3.44}$ as $r\to\infty$; see Figure \ref{fig:hires}.  At the
resolution limit of the \citeauthor{DiemandEtal2005} simulation (about 0.001 of
the virial radius), the fit has $\gamma\simeq1$, significantly larger than the
asymptotic cusp slope and still consistent with both the simulation data and the
behaviour of other fitting functions. Within the resolved radial range, our
model fit is also very closely traced by the best fit of the ad hoc profile
proposed by \cite{NavarroEtal2004}.

Our findings suggest a possible first-principles explanation for halo density
profiles along the following line of arguments. The initial distribution
function before collapse was completely cold (a $\delta$-function in velocity
space) and hence the phase-space density scale-free. The collapse and the
subsequent process of violent relaxation (phase-space mixing) is driven by
gravity alone, which cannot introduce any scale dependence. This implies that
the phase-space density of the collapsed halo satisfies some form of scale
invariance, suggesting that the ratio $\rho(r)/\sigma^3$, which is closely
related to the phase-space density, follows a power law (the general
scale-invariant functional form).  However, as our analysis has shown, if
$\rho(r)/\sigma^3$ of dark-matter haloes is \emph{any} power of radius, then the
condition of equilibrium (and physically sensible density profiles) requires the
\emph{particular} power law $r^{-\alpha_{\mathrm{crit}}}$ and, simultaneously,
that the density profile follows that given above.

From our fits to simulated haloes in Section~\ref{sec:simhalo}, the
$\beta$-$\gamma$ relation seems not as tight or universal as the $\rho$-$\sigma$
relation. This is not very surprising, since different amounts of velocity
anisotropy are required to stabilise different spatial halo shapes, resulting in
some scatter between the $\beta$-$\gamma$ relations of different haloes. Since
only the central velocity anisotropy $\beta_0$ affects the value of
$\alpha_{\mathrm{crit}}$, this scatter has little (or no) influence on the the
$\rho$-$\sigma$ relation as long as $\beta_0$ is similar (or the same) for
different haloes. This explains why haloes of different spatial shape still have
very similar density profiles.  Whether $\beta_0$ is a universal parameter
(close to zero) or whether there is some real scatter we cannot predict, but in
Section~\ref{sec:simhalo} we obtained a good fit to ten different simulated
haloes using a single value for $\beta_0$ (see Fig.~\ref{fig:simhalo}).

Of course, our analysis is still restricted to spherical symmetry and ignores
any halo substructure. However, substructure is unimportant for the issue of the
overall density profile, as simulations of dark-matter structure formation with
suppressed small-scale power in the initial conditions still yield the same
characteristic density profiles, but much less substructure
\citep{MooreEtal1999}. The issue of asphericity is more likely to be relevant
and hence it is all the more remarkable that our spherical analysis gives such a
good description of the spherically averaged profiles. Evidently, anisotropy,
which was ignored in previous studies, is presumably more important than
asphericity, because the gravitational potential is always less aspherical than
the density distribution.

\section*{acknowledgement}
It is a pleasure to thank J\"urg Diemand for kindly and promptly providing us,
in electronic form, with the density and kinematic radial profiles for simulated
CDM haloes. DEM is supported by a PPARC standard grant.  Research in theoretical
astrophysics at the University of Leicester is also supported by a PPARC rolling
grant.


\begin{thebibliography}{}

\bibitem[\protect\citeauthoryear{{Ascasibar}, {Yepes}, {Gottl{\" o}ber} \&
  {M{\" u}ller}}{{Ascasibar} et~al.}{2004}]{AscasibarEtal2004}
{Ascasibar} Y.,  {Yepes} G.,  {Gottl{\" o}ber} S.,    {M{\" u}ller} V.,  2004,
  MNRAS, 352, 1109

\bibitem[\protect\citeauthoryear{{Bertschinger}}{{Bertschinger}}{1985}]{Bertsc%
hinger1985}
{Bertschinger} E.,  1985, ApJS, 58, 39

\bibitem[\protect\citeauthoryear{{Bullock}, {Kolatt}, {Sigad}, {Somerville},
  {Kravtsov}, {Klypin}, {Primack} \& {Dekel}}{{Bullock}
  et~al.}{2001}]{BullockEtal2001a}
{Bullock} J.~S.,  {Kolatt} T.~S.,  {Sigad} Y.,  {Somerville} R.~S.,  {Kravtsov}
  A.~V.,  {Klypin} A.~A.,  {Primack} J.~R.,    {Dekel} A.,  2001, MNRAS, 321,
  559

\bibitem[\protect\citeauthoryear{{Carlberg}, {Yee}, {Ellingson}, {Morris},
  {Abraham}, {Gravel}, {Pritchet}, {Smecker-Hane}, {Hartwick}, {Hesser},
  {Hutchings} \& {Oke}}{{Carlberg} et~al.}{1997}]{CarlbergEtal1997}
{Carlberg} R.~G.,  {Yee} H.~K.~C.,  {Ellingson} E.,  {Morris} S.~L.,  {Abraham}
  R.,  {Gravel} P.,  {Pritchet} C.~J.,  {Smecker-Hane} T.,  {Hartwick}
  F.~D.~A.,  {Hesser} J.~E.,  {Hutchings} J.~B.,    {Oke} J.~B.,  1997, ApJ,
  485, L13

\bibitem[\protect\citeauthoryear{{Cole} \& {Lacey}}{{Cole} \&
  {Lacey}}{1996}]{ColeLacey1996}
{Cole} S.,  {Lacey} C.,  1996, MNRAS, 281, 716

\bibitem[\protect\citeauthoryear{{Col{\'{\i}}n}, {Klypin} \&
  {Kravtsov}}{{Col{\'{\i}}n} et~al.}{2000}]{ColinKlypinKravtsov2000}
{Col{\'{\i}}n} P.,  {Klypin} A.~A.,    {Kravtsov} A.~V.,  2000, ApJ, 539, 561

\bibitem[\protect\citeauthoryear{{Crone}, {Evrard} \& {Richstone}}{{Crone}
  et~al.}{1994}]{CroneEvrardRichstone1994}
{Crone} M.~M.,  {Evrard} A.~E.,    {Richstone} D.~O.,  1994, ApJ, 434, 402

\bibitem[\protect\citeauthoryear{{Diemand}, {Moore} \& {Stadel}}{{Diemand}
  et~al.}{2004a}]{DiemandMooreStadel2004a}
{Diemand} J.,  {Moore} B.,    {Stadel} J.,  2004a, MNRAS, 352, 535

\bibitem[\protect\citeauthoryear{{Diemand}, {Moore} \& {Stadel}}{{Diemand}
  et~al.}{2004b}]{DiemandMooreStadel2004b}
{Diemand} J.,  {Moore} B.,    {Stadel} J.,  2004b, MNRAS, 353, 624

\bibitem[\protect\citeauthoryear{{Diemand}, {Zemp}, {Moore}, {Stadel} \&
  {Carollo}}{{Diemand} et~al.}{2005}]{DiemandEtal2005}
{Diemand} J.,  {Zemp} M.,  {Moore} B.,  {Stadel} J.,    {Carollo} M.,  2005,
  MNRAS, submitted (astro-ph/0504215)

\bibitem[\protect\citeauthoryear{{Dubinski} \& {Carlberg}}{{Dubinski} \&
  {Carlberg}}{1991}]{DubinskiCarlberg1991}
{Dubinski} J.,  {Carlberg} R.~G.,  1991, ApJ, 378, 496

\bibitem[\protect\citeauthoryear{{Fukushige}, {Kawai} \& {Makino}}{{Fukushige}
  et~al.}{2004}]{FukushigeKawaiMakino2004}
{Fukushige} T.,  {Kawai} A.,    {Makino} J.,  2004, ApJ, 606, 625

\bibitem[\protect\citeauthoryear{{Fukushige} \& {Makino}}{{Fukushige} \&
  {Makino}}{1997}]{FukushigeMakino1997}
{Fukushige} T.,  {Makino} J.,  1997, ApJ, 477, L9

\bibitem[\protect\citeauthoryear{{Fukushige} \& {Makino}}{{Fukushige} \&
  {Makino}}{2001}]{FukushigeMakino2001}
{Fukushige} T.,  {Makino} J.,  2001, ApJ, 557, 533

\bibitem[\protect\citeauthoryear{{Ghigna}, {Moore}, {Governato}, {Lake},
  {Quinn} \& {Stadel}}{{Ghigna} et~al.}{1998}]{GhignaEtal1998}
{Ghigna} S.,  {Moore} B.,  {Governato} F.,  {Lake} G.,  {Quinn} T.,    {Stadel}
  J.,  1998, MNRAS, 300, 146

\bibitem[\protect\citeauthoryear{{Ghigna}, {Moore}, {Governato}, {Lake},
  {Quinn} \& {Stadel}}{{Ghigna} et~al.}{2000}]{GhignaEtal2000}
{Ghigna} S.,  {Moore} B.,  {Governato} F.,  {Lake} G.,  {Quinn} T.,    {Stadel}
  J.,  2000, ApJ, 544, 616

\bibitem[\protect\citeauthoryear{{Hansen} \& {Moore}}{{Hansen} \&
  {Moore}}{2005}]{HansenMoore2005}
{Hansen} S.,  {Moore} B.,  2005, MNRAS, submitted (astro-ph/0411473)

\bibitem[\protect\citeauthoryear{{Hansen}}{{Hansen}}{2004}]{Hansen2004}
{Hansen} S.~H.,  2004, MNRAS, 352, L41

\bibitem[\protect\citeauthoryear{{Hayashi}, {Navarro}, {Power}, {Jenkins},
  {Frenk}, {White}, {Springel}, {Stadel} \& {Quinn}}{{Hayashi}
  et~al.}{2004}]{HayashiEtal2004}
{Hayashi} E.,  {Navarro} J.~F.,  {Power} C.,  {Jenkins} A.,  {Frenk} C.~S.,
  {White} S.~D.~M.,  {Springel} V.,  {Stadel} J.,    {Quinn} T.~R.,  2004,
  MNRAS, 355, 794

\bibitem[\protect\citeauthoryear{{Hernquist}}{{Hernquist}}{1990}]{Hernquist199%
0}
{Hernquist} L.,  1990, ApJ, 356, 359

\bibitem[\protect\citeauthoryear{{Merritt}, {Navarro}, {Ludlow} \&
  {Jenkins}}{{Merritt} et~al.}{2005}]{MerrittEtal2005}
{Merritt} D.,  {Navarro} J.~F.,  {Ludlow} A.,    {Jenkins} A.,  2005, ApJ,
  624, L85

\bibitem[\protect\citeauthoryear{{Moore}, {Governato}, {Quinn}, {Stadel} \&
  {Lake}}{{Moore} et~al.}{1998}]{MooreEtal1998}
{Moore} B.,  {Governato} F.,  {Quinn} T.,  {Stadel} J.,    {Lake} G.,  1998,
  ApJ, 499, L5

\bibitem[\protect\citeauthoryear{{Moore}, {Quinn}, {Governato}, {Stadel} \&
  {Lake}}{{Moore} et~al.}{1999}]{MooreEtal1999}
{Moore} B.,  {Quinn} T.,  {Governato} F.,  {Stadel} J.,    {Lake} G.,  1999,
  MNRAS, 310, 1147

\bibitem[\protect\citeauthoryear{{Navarro}, {Frenk} \& {White}}{{Navarro}
  et~al.}{1996}]{NavarroFrenkWhite1996}
{Navarro} J.~F.,  {Frenk} C.~S.,    {White} S.~D.~M.,  1996, ApJ, 462, 563

\bibitem[\protect\citeauthoryear{{Navarro}, {Frenk} \& {White}}{{Navarro}
  et~al.}{1997}]{NavarroFrenkWhite1997}
{Navarro} J.~F.,  {Frenk} C.~S.,    {White} S.~D.~M.,  1997, ApJ, 490, 493

\bibitem[\protect\citeauthoryear{{Navarro}, {Hayashi}, {Power}, {Jenkins},
  {Frenk}, {White}, {Springel}, {Stadel} \& {Quinn}}{{Navarro}
  et~al.}{2004}]{NavarroEtal2004}
{Navarro} J.~F.,  {Hayashi} E.,  {Power} C.,  {Jenkins} A.~R.,  {Frenk} C.~S.,
  {White} S.~D.~M.,  {Springel} V.,  {Stadel} J.,    {Quinn} T.~R.,  2004,
  MNRAS, 349, 1039

\bibitem[\protect\citeauthoryear{{Plummer}}{{Plummer}}{1911}]{Plummer1911}
{Plummer} H.~C.,  1911, MNRAS, 71, 460

\bibitem[\protect\citeauthoryear{{Power}, {Navarro}, {Jenkins}, {Frenk},
  {White}, {Springel}, {Stadel} \& {Quinn}}{{Power}
  et~al.}{2003}]{PowerEtal2003}
{Power} C.,  {Navarro} J.~F.,  {Jenkins} A.,  {Frenk} C.~S.,  {White} S.~D.~M.,
   {Springel} V.,  {Stadel} J.,    {Quinn} T.,  2003, MNRAS, 338, 14

\bibitem[\protect\citeauthoryear{{Press}, {Teukolsky}, {Vetterling} \&
  {Flannery}}{{Press} et~al.}{1992}]{PressEtal1992}
{Press} W.~H.,  {Teukolsky} S.~A.,  {Vetterling} W.~T.,    {Flannery} B.~P.,
  1992, {Numerical Recipies in C}, 2nd edn.
Cambridge, Cambridge University Press

\bibitem[\protect\citeauthoryear{{Rasia}, {Tormen} \& {Moscardini}}{{Rasia}
  et~al.}{2004}]{RasiaTormenMoscardini2004}
{Rasia} E.,  {Tormen} G.,    {Moscardini} L.,  2004, MNRAS, 351, 237

\bibitem[\protect\citeauthoryear{{Stoehr}, {White}, {Tormen} \&
  {Springel}}{{Stoehr} et~al.}{2002}]{StoehrEtal2002}
{Stoehr} F.,  {White} S.~D.~M.,  {Tormen} G.,    {Springel} V.,  2002, MNRAS,
  335, L84

\bibitem[\protect\citeauthoryear{{Taylor} \& {Navarro}}{{Taylor} \&
  {Navarro}}{2001}]{TaylorNavarro2001}
{Taylor} J.~E.,  {Navarro} J.~F.,  2001, ApJ, 563, 483

\bibitem[\protect\citeauthoryear{{Williams}, {Austin}, {Barnes}, {Babul} \&
  {Dalcanton}}{{Williams} et~al.}{2004}]{WilliamsEtal2004}
{Williams} L.~L.~R.,  {Austin} C.,  {Barnes} E.,  {Babul} A.,    {Dalcanton}
  J.,  2004, in {Dettmar} R.,  {Klein} U.,   {Salucci} P.,  eds, {Baryons in
  Dark Matter Halos} Proceedings of Science.
pp 20--23

\bibitem[\protect\citeauthoryear{{Zhao}}{{Zhao}}{1996}]{Zhao1996}
{Zhao} H.,  1996, MNRAS, 278, 488

\end{thebibliography}

\label{lastpage}
\end{document}